# Cosmology, astrobiology, and the RNA world. Just add quintessential water

Keith Johnson
Massachusetts Institute of Technology, Cambridge, MA 02139

## Abstract

Laboratory generation of water nanoclusters from amorphous ice and strong terahertz (THz) radiation from water nanoclusters ejected from water vapor into a vacuum suggest the possibility of water nanoclusters ejected into interstellar space from abundant amorphous ice-coated cosmic dust produced by supernovae explosions. Water nanoclusters (Section 2) offer a hypothetical scenario connecting major mysteries of our universe: dark matter (Section 3), dark energy (Section 4), cosmology (Section 5), astrobiology (Section 6), and the RNA world (Section 7) as the origin of life on Earth and habitable exoplanets. Despite their expected low density in space compared to hydrogen, their quantum-entangled diffuse Rydberg electronic states make cosmic water nanoclusters a candidate for baryonic dark matter that can also absorb, via the microscopic dynamical Casimir effect, the virtual photons of zero-point-energy vacuum fluctuations above the nanocluster cut-off vibrational frequencies, leaving only vacuum fluctuations below these frequencies to be gravitationally active, thus leading to a possible common origin of dark matter and dark energy. This picture includes novel explanations of the small cosmological constant, the coincidence of energy and matter densities, possible contributions of the red-shifted THz radiation from cosmic water nanoclusters at redshift $z \cong 10$ to the CMB spectrum, the Hubble constant crisis, the role of water as a known coolant for rapid early star formation, and ultimately how life may have originated from RNA protocells on Earth and exoplanets and moons in the habitable zones of developed solar systems. Together they lead to a cyclic universe cosmology - based on the proposed equivalence of cosmic water nanoclusters to a quintessence scalar field - instead of a multiverse based on cosmic inflation theory. Cosmic water nanoclusters also exhibit the dipole-moment anisotropy prerequisite to their birefringence property, which may explain recent CMB birefringence data possibly supporting quintessence. Finally, from the quantum chemistry of water nanoclusters interacting with prebiotic organic molecules, amino acids, and RNA protocells on early Earth and habitable exoplanets, this scenario is consistent with the anthropic principle that our universe must have properties which allow life, as we know it - based on water, to develop at the present stage of its history.

## 1. Introduction

This paper is a 'gedanken experiment'. Many of the following ideas need to be tested by future experimental observations, and some are speculative, requiring further discussion and more formal mathematical models.

In the standard model of Big Bang cosmology, the matter and energy resources of our universe are controlled respectively by *dark matter* (Weinberg 2008) - a *nonbaryonic* substance of undetermined nature - and *dark energy* (Riess et al. 1998) - a negative-pressure field of exotic physical origin. Over thirty years of searching for dark matter, proposed exotic elementary particles such as *weakly interacting massive particles* (WIMPS) and AXIONS thus far have not been observed experimentally, even in the latest Large Underground Xenon (LUX) and MIT ABRACADABRA detectors, respectively (Akerib et al. 2017; Ouellet et al. 2019). Nor have the WIMPS predicted from supersymmetry theory been created in the CERN LHC or ATLAS detector. The dark energy believed to be responsible for the accelerating expansion of our universe is usually considered to be a separate problem from dark matter and to be associated with zero-point-energy fluctuations of the cosmic vacuum. However, quantum field theory predicts a vacuum energy density that is too large by up to a factor of $10^{120}$ (Zeldovich and Krasinski. 1968; Weinberg 1989), which is the well-known cosmological constant problem. I propose that nanoclusters of water molecules ejected by cosmic rays from amorphous ice layers on ubiquitous cosmic dust produced from exploding supernovae (Matsuura et al. 2019), albeit at low density compared to elemental hydrogen and oxygen, excited to their diffuse *Rydberg states* (Herzberg 1987), are a possible candidate for *baryonic* dark matter. The cut-off terahertz (THz) vibrational frequencies of such water nanoclusters are close to the $\nu_c \cong 1.7$ THz cut-off frequency of zero-point-energy vacuum fluctuations proposed to account for the small value of vacuum energy and cosmological constant (Beck and Mackey 2005, 2007). Beck and Mackey (2005) proposed laboratory measurements of dark energy, which were challenged by Jetzer and Straumann (2005, 2006), to which Beck and Mackey (2006) rebutted. In this paper, cosmic water nanoclusters are postulated to capture via the *microscopic dynamical Casimir effect* (Souza, Impens, and Neto 2018) the high-frequency vacuum zero-point-energy virtual photons, leaving only the low-frequency ones to be gravitationally active. Water constitutes approximately seventy percent of our body weight, much of it as water nanoclusters called "structured" water (Chaplin 2006; Johnson 2012). This fact adds scientific and philosophical support based on the simplest *anthropic principle* (Weinberg 1987) to the proposal that water nanoclusters distributed as a low-density "dark fluid" throughout our universe are a possible common origin of dark matter and dark energy along with or instead of thus-far undiscovered exotic elementary particles.

Section 2 of the paper covers water-nanocluster electronic structure and vibrations. Sections 3 - 7 describe their possible relevance to dark matter, dark energy, cosmology, astrobiology, and the RNA world. Section 8.4 of the Conclusions and Fig.10 deal with birefringence, supplementing the publication (Johnson 2021).

## 2. Water nanoclusters

### 2.1 Cosmic implications

Since hypothetical WIMP and AXION elementary particles, after many years of costly experiments, have still not been found, and *baryonic* dark matter has been largely ruled out by standard cosmological theory, why would one expect water nanoclusters to exist in the cosmos, and why would they have anything to do with dark matter and/or dark energy? Hydrogen and oxygen are the most abundant chemically reactive elements in our universe, oxygen slightly beating carbon. Water vapor plays a key role in the early stages of star formation, where it is an important oxygen reservoir in the warm environments of star-forming regions, and is believed to contribute significantly to the cooling of the circumstellar gas, thereby removing the excess energy built up during proto-stellar collapse (Bergin and van Dishoeck 2012). For example, the star-forming region of the Orion nebula produces enough water in a day to fill up earth's oceans many times over (Glanz 1998). The largest and farthest reservoir of water ever detected in the universe has been reported to exist in a high-redshift (z = 3.91) quasar approximately 12 billion light-years away (Bradford et al. 2011). The quasar water vapor mass is at least 140 trillion times that of all the water in the world's oceans and 100,000 times more massive than the sun. It has been proposed recently that water could have been abundant during the first billion years after the Big Bang (Bialy, Sternberg, and Loeb 2015). In Section 5.2 of this paper, the possible role of water nanoclusters as a coolant catalyst for rapid early star formation in high-redshift clouds is discussed.

On planet Earth, through hydrogen bonding between water monomers, stable water nanoclusters, both neutral and ionized, are easily formed in molecular beams (Carlon 1981), occur naturally in the water vapor of earth's atmosphere (Aplin and McPheat 2005), and are produced from amorphous ice by energetic ion bombardment (Martinez et al. 2019). In the cosmos, therefore, a natural route to water nanocluster formation would be via the ejection from amorphous water-ice coatings of cosmic dust grains (Dulieu et al. 2010; Potapov, Jager, and Henning 2020), which are believed to be abundant in interstellar clouds because they are a product of supernovae explosions (Matsuura et al. 2019). As a prime example, cosmic ray ionization of $H_2$ molecules adsorbed on amorphous ice-coated dust grains can lead to the reaction (Duley 1996)

$$H_2^+ + nH_2O + \text{grain} \rightarrow H_3O^+(H_2O)_{n-1}\uparrow + \text{grain}. \quad (1)$$

Interstellar population of protonated water-nanocluster ions released by this process has been estimated to approach $10^{-6}$ of the average atomic hydrogen population but could likely be significantly greater (Martinez et al. 2019).  Due to their large electric dipole moments ($\geq$ 10D) oscillating at THz frequencies, such water nanoclusters are believed to be responsible for the observed strong THz emission from water vapor into a vacuum under intense U.V. optical stimulation (Johnson et al. 2008) (Fig. 1) and therefore should be relatively stable under similar cosmic radiation.

While $H_3O^+(H_2O)_{n-1}$ nanoclusters ejected by ion bombardment from amorphous ice have been observed over a range of n-values (Martinez et al. 2019), the "magic-number" n = 21 pentagonal dodecahedral $H_3O^+(H_2O)_{20}$ or equivalent protonated $(H_2O)_{21}H^+$ nanocluster (Fig. 2) is exceptionally stable in a vacuum and is of potential cosmic importance because it can be viewed as a $H_3O^+$ (hydronium) ion caged ("clathrated") by an approximately pentagonal dodecahedral cage of twenty water molecules (Miyazaki et al. 2004; Shin et al. 2004).  Interstellar $H_3O^+$ has been observed recently (Lis et al. 2014) and its discovery points to the challenge of trying to identify spectroscopically larger cosmic water nanoclusters such as $H_3O^+(H_2O)_{20}$ because both spectra fall into the same THz region. The occurrence of stable pentagonal dodecahedral water nanocluster clathrate hydrates in the interstellar medium has recently been predicted (Ghosh, Methikkalam, and Bhuin 2019). References (Miyazaki et al 2004; Shin et al. 2004; Lis et al. 2014) could be the starting point for attempts to identify the cosmic presence of $H_3O^+(H_2O)_{20}$.

## 2.2 Electronic structure and THz vibrations

Fig. 2 shows the ground-state molecular-orbital energies, wavefunctions, and vibrational modes of the pentagonal dodecahedral $(H_2O)_{21}H^+$ or $H_3O^+(H_2O)_{20}$ cluster computed by the SCF-X$\alpha$-Scattered-Wave density-functional method co-developed by the author (Slater and Johnson 1972, 1974).  Molecular dynamics simulations yield results qualitatively unchanged at temperatures well above 100C, where the cluster remains remarkably intact. Similar calculations for the neutral pentagonal dodecahedral water cluster, $(H_2O)_{20}$ and arrays thereof have also been performed, yielding the THz vibrational modes displayed in Figs. 3 and 4.

These results are qualitatively similar to those shown in Fig. 2, but they indicate a gradual decrease of the cluster cut-off vibrational frequency with increasing cluster size. The latter trend correlates with the experimental studies of THz radiation emission from water vapor nanoclusters (Johnson et al. 2008), showing in Fig. 1 the shift in the cluster THz emission peaks toward lower frequencies and intensities – corresponding to a trend toward larger clusters – with decreasing vapor ejection pressure into the vacuum chamber where the radiation was measured.  Relating this finding to Eq. 1 would suggest decreasing THz emission cut-off frequencies and intensities with the increasing sizes (increasing n-values) of water nanoclusters ejected from ice-coated cosmic dust.

Common to all these water clusters are: (1) lowest unoccupied (LUMO) energy levels like those in Fig. 2a, which correspond to the diffuse Rydberg "S"-, "P"-, "D"- and "F"-like cluster "surface" molecular-orbital wavefunctions shown in Fig. 2b, and (2) bands of vibrational modes between 0.5 and 6 THz (Figs 2 – 4), due to O-O-O "squashing" (or "bending") and "twisting" motions between adjacent hydrogen bonds.  The vectors in Figs. 2 – 4 represent the directions and relative amplitudes of the lowest THz-frequency modes corresponding to the O-O-O "bending" (or "squashing") motions of the water-cluster "surface" oxygen ions.  Surface O-O-O bending vibrations of water clusters in this energy range have indeed been observed under laboratory conditions (Brudermann, Lohbrandt, and Buck 1998). Ultraviolet excitation of an electron from the HOMO to LUMO (Fig 2a) can put the electron into the Rydberg "S"-like cluster molecular orbital mapped in Fig 2b.  Occupation of this orbital produces a bound state, even when an extra electron is added, the so-called "hydrated electron" (Jordan 2004).  In contrast, a water monomer or dimer has virtually no electron affinity. Therefore, in space – especially within dense interstellar clouds – $(H_2O)_{21}H^+$ or $H_3O^+(H_2O)_{20}$ and larger water-nanocluster ions ejected from ice-coated cosmic dust according to Eq. 1 are likely to capture electrons, forming electrically neutral water nanoclusters of the types shown in Fig. 3.

## 3. Baryonic dark matter

### 3.1 Rydberg matter

Stellar electromagnetic radiation can potentially stimulate electronic excitations from the HOMO of $(H_2O)_{21}H^+$ (Fig. 2a) (or from the LUMO with a captured hydrated electron) to the increasingly diffuse "P", "D", "F" and higher water cluster Rydberg orbitals in Fig. 2b (Herzberg 1987; Holmlid 2008). These states have vanishing spatial overlap with the lower-energy occupied ones, have long lifetimes that increase with increasing excitation energy and effective principal quantum number, and thus are candidates for *Rydberg Matter* (RM) – a *low-density* condensed phase of weakly interacting individual Rydberg-excited molecules with long-range effective interactions (Badiei and Homlid 2002). RM can interact or become *quantum-entangled* over long effective distances, causing it to be transparent to visible, infrared, and radio frequencies, and thus qualifies as *baryonic dark matter* (Badiei and Homlid 2002). Two water nanoclusters of the forms shown in Figs. 2 and 3a, but each holding an excited or "hydrated" electron in the Rydberg "S" LUMO, are like giant hydrogen atoms, which for short distances between the clusters will form an overlapping "bonding" molecular orbitals holding two spin-paired electrons in analogy to a giant hydrogen molecule. However, the approach of two water nanoclusters to each other in interstellar space should be a rare occurrence because of their relatively low density. For much larger distances between the clusters, the diffuse molecular orbitals of their highest-energy Rydberg states "overlap" sufficiently to permit quantum entanglement of water nanoclusters over long distances in space, thus qualifying cosmic water nanoclusters as possible baryonic dark matter. These nanoclusters can be interpreted as a *scalar field* permeating space – a type of *quintessence* (Ratra and Peebles 1988; Steinhardt 2003) leading to a time-dependent dark energy density (see Section 4).

### 3.2 The Bullet cluster and galactic halos

Can low-density, quantum-entangled water-nanocluster RM account for at least part of the dark matter estimated from inflationary Big Bang theory? The consensus of standard cosmology is that the unknown dark matter cannot be baryonic. Gravitational lensing observations of the *Bullet Cluster* revealing the separation of normal luminous matter and dark matter have been said to be the best evidence to date for the existence of nonbaryonic dark matter (Clowe, Gonzalez, and Markevich 2004). Protonated water nanoclusters produced by Eq. 1 and shown in Fig. 2 for n = 21 are positively charged, although, as pointed out above, such clusters are likely to pick up a "hydrated" electron (Jordan 2004) from space once ejected from ice-coated cosmic dust, forming electrically neutral clusters like those pictured in Fig. 3. It is believed that nonbaryonic dark matter is uncharged. Nevertheless, it has recently been argued that a small amount of charged dark matter could cool the baryons in the early universe (Munoz and Loeb 2018). While magnetic fields associated with celestial objects should interact with water nanocluster ions, which are motional sources of magnetism, the origin and relevance of intergalactic magnetic fields is still debated (Jedamzik and Pogosian 2020). Magnetic fields may be important to the possible role of water nanoclusters in water vapor as a coolant catalyst for rapid early star formation, discussed in Section 5.2 of this paper. The electrical charge of the ice-coated cosmic dust that is the postulated origin of cosmic water nanoclusters according to Eq. 1 may be key to the properties of the Bullet Cluster dark matter (Clowe, Gonzalez, and Markevich 2004). Because much of the Bullet Cluster normal matter is likely composed of positively charged cosmic dust (Mann 2001), its electrical repulsion of protonated water nanoclusters would enhance the ejection of water nanoclusters described by Eq. 1, thus explaining the observed separation of normal luminous matter and dark matter. Despite uncertainties about electric fields on the galactic scale (Bally and Harrison 1978; Chakraborty et al. 2014), it is possible that such fields could cause cosmic water nanocluster RM to aggregate around the peripheries of galaxies, thereby possibly explaining *galactic dark matter halos* similarly to the Bullet Cluster dark matter.

## 4. Dark energy

Quantum field theory predicts a vacuum energy density that is too large by a factor of $10^{120}$, which leads to the well-known cosmological constant problem (Weinberg 1989). It has been suggested that if the observed dark energy responsible for the accelerated expansion of the universe is equated to the otherwise infinite cosmic vacuum energy density predicted by theory, then gravitationally active zero-point-energy vacuum fluctuations must have a cut-off frequency of $\nu_c \cong 1.7$ THz (Beck and Mackey 2005, 2007). In other words, the virtual photons associated with vacuum fluctuations should be of gravitational significance only below this frequency to be consistent with the observationally small magnitude of dark energy. A $\nu_c \cong 1.7$ THz vacuum fluctuation cut-off frequency is the same order of magnitude as the cut-off vibrational frequencies of prominent water nanoclusters, although these frequencies decrease with increasing cluster size (Figs. 2 - 4) or increasing n-value in Eq. 1. Other molecules in space, such as hydrogen, water monomers, carbon buckyballs, and various observed organic molecules do not have vibrational cut-off frequencies in this THz region. As low-density Rydberg matter (Section 3.1), cosmic water nanoclusters can be viewed as constituting a *quintessence scalar field* Q of energy density, $\rho = ½\dot{Q}^2 + V(Q)$ (Steinhardt 2003), whose properties depend on the absorption of the virtual photons of vacuum fluctuations at frequencies greater than $\nu_c \cong 1.7$ THz via the *microscopic dynamical Casimir effect* (Souza, Impens, and Neto 2018; Leonhardt 2020). This process converts the virtual photons to real ones, leaving only the vacuum low-frequency photons to be gravitationally active (Fig. 5). The absorbed photons can then decay via emitted THz radiation (Fig. 1).

To quantify this, we conventionally view the vacuum electromagnetic field (excluding other fields) as a collection of harmonic oscillators of normal-mode frequencies $\nu_k$, summing over the zero-point energies of each oscillator mode, leading to the following energy density

$$\rho_{vac} = \frac{E}{V} = \frac{1}{V} \sum_k \frac{1}{2} h\nu_k = \frac{4\pi h}{c^3} \int_0^\infty \nu^3 \, d\nu \qquad (2)$$

where the wave vector **k** signifies the normal modes of the electromagnetic field that are consistent with the boundary conditions on the quantization volume *V*. As *V* approaches infinity, one obtains the right-hand side of Eq. 2. The divergent integral in Eq. 2 can be avoided by replacing the upper limit by a cut-off frequency $\nu_c$ set by the Planck scale (Weinberg 1989). However, this results in a huge vacuum energy that exceeds the cosmologically measured value by 120 orders of magnitude. If instead we subtract from (2) the energy density

$$\rho_c = \frac{4\pi h}{c^3} \int_{\nu_c}^\infty \nu^3 \, d\nu \qquad (3)$$

of the virtual photons of zero-point vacuum fluctuations captured by the water clusters through the microscopic dynamical Casimir effect (Souza, Impens, and Neto 2018), the divergent integral in Eq. 2 is largely cancelled, leaving the finite quantity, Eq. 4 to be identified with the dark energy density

$$\frac{4\pi h}{c^3} \int_0^{\nu_c} \nu^3 \, d\nu = \frac{\pi h \nu_c^4}{c^3} = \rho_{dark}. \qquad (4)$$

Due to the small nanocluster vibrational kinetic energies $½\dot{Q}^2$ compared to their potential energy $V(Q)$, which is elevated to higher-THz-frequency "surface" vibrational modes (Figs. 2-4) by the capture of vacuum photons, as shown schematically in Fig. 5, it follows that the *quintessence scalar field pressure*, $P = ½\dot{Q}^2 - V(Q)$ (Steinhardt 2003) becomes more negative with increasing $\nu_c$ and therefore with $\rho_{dark}$. The PLANCK observations have concluded that dark energy presently constitutes 68.3% of the total known energy of the universe (Ade et al.), leading to $\rho_{dark} = 3.64$ GeV/m³. Eq. 4 then requires a cut-off frequency of $\nu_c = 1.66$ THz, which is the same order of magnitude as the cut-off frequencies of the smallest pentagonal dodecahedral water clusters shown in Figs. 2d and 3a. However, since *$\nu_c$ decreases with increasing water-cluster size* (Figs. 3 and 4) or with increasing n-value in Eq. 1, a trend toward the ejection of larger water clusters from cosmic dust over time would imply a *decrease of dark energy density over time* according to Eq. 4, and therefore a *decreasing acceleration of the universe*.

## 5. Cosmology

### 5.1 The CMB spectrum

The consensus of standard inflationary cosmology (Guth 1981, 2007; Linde 2008) is that the measured cosmic microwave background (CMB) spectrum of the universe has its origin at approximately 380 thousand years after the Big Bang (Ade et al. 2016; Bennett et al. 2013). Is there a possible and credible additional contribution to the CMB that is consistent with its spectrum and the THz vibrational properties of water nanoclusters discussed in the previous sections? It was suggested long ago that the CMB might be attributable to thermalization by "cosmic dust" in the form of hollow, spherical shells of high dielectric constant or conducting "needle-shaped grains" (Layzer and Hively 1973; Wright 1982). Layzer and Hively (1973) argued that a relatively low density of high-dielectric-constant dust could thermalize the radiation produced by objects of galactic mass at redshift $z \cong 10$. Because of their computed large electric dipole moments and measured strong THz radiation emission (Johnson et al. 2008) (Fig. 1), optically pumped water nanoclusters in water vapor consisting of spherical "shells" of water-cluster O-H bonds (Figs. 2 and 3) or "strings" of water clusters (Fig. 4) satisfy the conditions proposed in (Layzer and Hively 1973; Wright 1982). The thin amorphous water ice that coats the cosmic dust from which water clusters are ejected according to Eq. 1 can be viewed as disordered water nanoclusters of high dielectric constant and could directly contribute. Astronomical observations have pushed back the epoch of protogalaxy formation and reionization to redshifts of $z = 8.6$ and $z = 9.6$, respectively, *i.e.* to corresponding times of 600 and 500 million years after the Big Bang (Lehnert et al. 2010; Zheng et al. 2012), although more recently, the Hubble Space Telescope has found a galaxy at $z \cong 11$, corresponding to just 400 million years after the Big Bang (Oesch et al. 2016). At redshift $z \cong 10$, the distinctive THz vibrational manifolds of water clusters (Figs. 2 - 4), as well as the laboratory THz emission peaks of Fig. 1 are red-shifted to the region of the measured CMB spectrum, suggesting water nanocluster THz emission originating at $z \cong 10$, where the temperature $T \cong 30K$ is compatible with the existence of such nanoclusters, could indeed contribute to the CMB in addition to photons from the "recombination" period at $z \cong 1100$, where the temperature is $T \cong 4000K$.

Laboratory measurements (Johnson et al. 2008) (Fig. 1) of the THz emission from water nanoclusters as a function of ejection pressure into a vacuum chamber indicate emission peaks that decrease in frequency and intensity with decreasing pressure and thus, according to Figs. 2-4, with increasing cluster size. The most intense emission peak at approximately 1.7 THz is assigned to water clusters of the "magic numbers" $n = 21$ and 20 shown in Figs. 2d and 3a, respectively, whereas the peaks decreasing in frequency and intensity with lower pressure to approximately 0.5 THz are due to larger clusters like those shown in Figs. 3b,c and 4a. The effective temperatures of the water nanoclusters scale with this pressure trend, with the smallest clusters emitting the "hottest" radiation peaking around 1.7 THz. Applying this power spectrum to the redshifted radiation from water nanoclusters ejected from ice-coated cosmic dust at $z \cong 10$, the spectrum of the smallest versus larger water nanoclusters in particular regions of space might be similar to the measured CMB power spectrum (Ade et al. 2016; Bennett et al. 2013), but is a subject for future investigation requiring more resources by this author. The laboratory ejection pressure dependence of the vacuum chamber power spectrum shown in Fig. 1 applied to the effective water-nanocluster ejection "pressures" from ice-coated cosmic dust according to Eq. 1 further suggests a "pressure wave" created at $z \cong 10$, or around 500 million years after the Big Bang, in analogy to the CMB acoustic wave assigned to the "recombination" period at $z \cong 1100$ or 380 thousand years after the Big Bang (Ade et al. 2016; Bennett et al. 2013). In other words, if one were to interpret simplistically the measured CMB power spectrum and its anisotropy as due at least partially to the redshifted THz radiation from water nanoclusters ejected from ice-coated cosmic dust at $z \cong 10$, this would suggest a relatively slow "classical" process extending over billions of years compared to an inflationary hot Big Bang originating from a quantum singularity. Adding to the possible credibility of this scenario is the fact that $z \cong 10$, or around 500 million years after the Big Bang, is near the first times of suspected early (Population III) star formation, with most of these thus far hypothetical stars having short lives and becoming explosive supernovae that produce the cosmic dust from which cosmic water nanoclusters can be ejected according to Eq. 1.

## 5.2 Early star formation

Water vapor is a recognized coolant for star formation (Bergin and Dishoeck 2012). The report of a huge reservoir of water in a high-redshift ($z \cong 4$) quasar, corresponding to a water vapor mass at least 140 trillion times that of all the water in the world's oceans and 100,000 times more massive than the sun (Bradford et al. 2011), together with the recent proposal that water vapor could have been abundant during the first billion years after the Big Bang (Bialy, Sternberg, and Loeb 2015) suggests the possibility of a significant population of stable water nanoclusters at redshift $z \cong 10$ or around 500 million years after the Big Bang.

The rapidness at which some early stars were created from observed dense gas clouds at $z \cong 6.4$ or 850 million years after the Big Bang (Banandos et al. 2019) can possibly be understood by the presence of water nanoclusters in the cloud's water vapor coolant. Studies of the infrared absorption by water nanoclusters in the laboratory and Earth's atmosphere (Carlon 1981; Aplin and McPheat 2005), including both protonated cluster ions of the type shown in Fig 2 and neutral clusters like those shown in Fig. 3, have established their extraordinary heat storage as due mainly to the *nanocluster librational modes* - especially those near 32 THz (1060 cm$^{-1}$) - shown in these figures. The cooling effect of the star-forming gas clouds can consequently be achieved through the release of photons associated with the *cluster surface modes* in the 1 – 6 THz range (Figs. 2 and 3). These photons are the same ones that might contribute to the CMB, as described above. In other words, if one accepts a scenario where redshifted THz radiation from cosmic water nanoclusters at $z \cong 10$ contributes to the CMB spectrum, then the CMB also possibly contains information about early star formation.

## 5.3 The Hubble constant crisis

The present model suggests a possible resolution of the Hubble constant "crisis," where the value of this constant deduced from observations of supernovae and cepheids has indicated the universe is expanding significantly faster than the value concluded from measurements of the microwave radiation emitted immediately after the Big Bang. The ice-coated cosmic dust responsible for ejecting the water nanoclusters proposed herein to underlie dark energy and our accelerating universe is a product of stellar evolution. Since the first stars were born only after the reionization phase following the recombination phase of hydrogen formation that began approximately 380,000 years after the Big Bang, there would be no significant cosmic dust and thus no cosmic water nanoclusters until many years after the Big Bang. This conclusion is supported by a recent observation of the oldest cosmic dust 200 million years after the birth of the first stars (Laporte et al. 2017). As the universe expanded over 13.8 billion years, the amount of cosmic dust increased with the formation of more stars and galaxies until one reached the present era where there is enough dust and ejected water-nanocluster Rydberg matter to account for the Hubble constant deduced from recent supernovae and cepheid observations, as well as the current coincidence of dark energy and matter densities. This scenario is consistent with studies of high-redshift quasars showing that dark energy has increased from the early universe to the present (Risaliti and Lusso 2019). It is also somewhat consistent with a claim that baryon inhomogeneities explain away the Hubble crisis but disagrees that they are due to primordial magnetic fields (Jedamzik and Pogosian 2020).

## 5.4 A cyclic cosmology

Inflationary cosmology (Guth 1981, 2007; Linde 2008) leads to the conclusion that dark matter density should decrease faster than dark energy in an accelerating universe, so that eventually dark energy will become dominant, and the ultimate fate of the universe is that all the matter in the universe will be progressively torn apart by its expansion – the so-called "big rip". In contrast, if we conceptually view cosmic water nanoclusters - ejected from ice-coated cosmic dust to form Rydberg matter - as equivalent to a time-dependent *quintessence scalar field* (Steinhardt 2003), as described in Section 4, the present model suggests the following hypothetical cyclic dark-matter-dark-energy cosmology:

(1) As the universe expands and stellar nuclear fusion produces heavier elements, more water-ice-coated cosmic dust will be produced from the increasing explosive supernovae population. With the expanding volume of space and decreasing pressure, the growing dust presence over time will expel larger water clusters according to Fig. 1. A trend to lower vibrational cut-off frequency $\nu_c$ with increasing water-cluster size (increasing n-value in Eq. 1) is suggested by the computed results shown in Figs. 2 - 4.

(2) With the increasing population of larger water clusters such as the one shown in Fig. 4 over astronomical time, there will be a trend to clusters having a much lower vibrational cut-off frequency approaching $\nu_c \cong 0.5$ THz (see also Fig. 1), resulting in a movement to a much lower dark energy density $\rho_{dark}$ approaching zero according to Eq. 4 over astronomical time and thus to a practically vanishing acceleration of the expanding universe. Possible time dependences of $\rho_{dark}$ in an expanding universe have been discussed by others (Linder and Jenkins 2003; Sola 2014).

(3) At that point in time – likely billions of years from now - the gravity of remaining baryonic mass, consisting mostly of ice-coated cosmic dust produced by a declining supernovae population, will take over. The universe will stop expanding, begin to collapse, and *slowly* return - without "crunching" the remaining baryonic matter - to the $z \cong 10$ period around 500 million years after the Big Bang, where the size of the universe will be approximately ten percent of the present one.

(4) As the latter takes place, pressure of the expelled water-nanocluster vapor expected to exist at $z \cong 10$ (Bialy, Sternberg, and Loeb 2015) will increase again and, according to Fig. 1, smaller water nanoclusters – especially those with the magic numbers n = 21,20 and higher cut-off vibrational frequencies $\nu_c$ shown in Figs. 2d and 3a, respectively - will again be favored over larger ones. Those water nanoclusters will be available once more as a coolant catalyst for Population III star formation, as described in Sections 5.1 and 5.2, evolving again to explosive supernovae that produce the cosmic dust from which cosmic water nanoclusters can be ejected according to Eq. 1.

(5) This is a possible *classical nonsingular* starting point for *non-inflationary re-expansion*, although further collapse toward the reionization period and decomposition of water vapor to supply hydrogen for star formation and non-inflationary re-expansion is also a likelihood. Multi-billion-year or longer cycle periods follow from the estimated dust-producing supernovae population cycles, although their magnitudes remain an open question to be addressed further. This cyclic universe model mimics ones proposed by others (Ijjas, Steinhardt, and Loeb 2017; Ijjas and Steinhardt 2019; Penrose 2006) and likewise lessens the credibility of the *multiverse theory* that is an evolutionary part of inflation theory.

(6) This scenario does not seem to violate the second law of thermodynamics and leads to the conclusion that we are likely living at the ideal time in the expansion of the universe for life - as we know it - to exist, as described in the following sections devoted to astrobiology and the RNA World.

## 6. Astrobiology

The discovery of organic molecules, including prebiotic ones, in interstellar space, dust clouds, comets, and meteorites over the past fifty years has been impressive (Kvenvolden et al. 1970; Lacy et al. 1991; Iglesias-Groth et al. 2010). While the significance of cosmic carbon compounds to the possible existence of life throughout the universe has most often been emphasized, it is ultimately water that is the key to the structures and functions of carbon-based biomolecules. The human body, which is approximately 70% water by weight, cannot exist without water, which is essential for the synthesis of RNA, DNA, and proteins. Much of this water is attached to proteins, RNA, and DNA as water nanoclusters, called "structured" water (Chaplin 2006; Johnson 2012). For example, proteins not containing nanostructured water will not fold properly and can lead to degenerative diseases such as Parkinson's, Alzheimer's, and cataracts, while such structured water is also essential to RNA and DNA replication. How water nanoclusters interact with organic molecules in astrobiology can be investigated by first-principles quantum chemistry calculations using the SCF-Xα-Scattered-Wave density-functional method (Slater and Johnson 1972, 1974).

This method was first applied to proteins (Cotton et al. 1973; Yang et al. 1975; Case, Huynh, and Karplus 1979). The resulting molecular structures and lowest THz-frequency vibrational modes of a pentagonal dodecahedral water nanocluster interacting with *methane, anthracene*, and the *amino acid valine*, respectively, all of which have been found in interstellar space, dust clouds, comets, or meteorites (Lacy et al. 1991; Iglesias-Groth et al. 2010; Kvenvolden et al. 1970), are shown in Fig. 6. In all three examples, there is coupling of 1.8 THz water-cluster "surface" oxygen motions to the carbon atomic motions, as represented by the vectors in Fig. 6. In the anthracene and valine examples there are carbon-carbon bonds, so that the water-cluster-induced carbon motions at 1.8 THz are "bending" modes of the C-C bonds. In valine, there is also coupling between the water-cluster 1.8 THz vibrational mode to that of the nitrogen atom (Fig, 6f). Valine is an α-amino acid that is used in the biosynthesis of proteins by polymerization beginning with the nitrogen atom and ending with a carbon. The point here is that water nanoclusters and prebiotic molecules delivered by cosmic dust and meteorites to Earth could have jump-started life here and on exoplanets in the habitable zones of distant solar systems. This requires the first self-replicating RNA necessary for DNA and the synthesis of proteins from amino acids.

## 7. The RNA World

### 7.1 From prebiotic molecules to RNA

Compelling arguments for the so-called *RNA World* as the origin of life on planet Earth and possibly elsewhere in the universe (Joyce and Orgel 1993) – preceding DNA- and protein-based life – has posed the fundamental problem of explaining how the first self-replicating RNA polymers, such as the RNA segment shown in Fig. 7, were created chemically from a pool of prebiotic organic molecules, nucleosides, and phosphates. A recent paper (Totani 2020) based on polymer physics and inflationary cosmology proposes that extraterrestrial RNA worlds and thus the emergence of life in an inflationary universe must be statistically rare. Since the polymerization of long RNA chains in water has been demonstrated but not explained definitively (Costanzo et al. 2009), it is proposed here that water nanoclusters comprising liquid water and water vapor can act as catalysts for prebiotic RNA synthesis, increasing the likelihood of RNA worlds and thus the emergence of life wherever water is present in a non-inflationary or cyclic universe of the type described in this paper.

Simply described, the chemical steps of combining the prebiotic molecules of Fig 7 to yield an RNA sequence of four polymerized nucleobases, *guanine*, *adenine*, *uracil*, and *cytosine*, require the effective loss of eleven water molecules from the initial reactants and their effective recombination in RNA. This is a *dehydration-condensation reaction* (Cafferty and Hud 2014). Cyclic water pentamers (Harker et al. 2005) (Fig. 7) have been identified as being key to the hydration and stabilization of biomolecules (Teeter 1984). Such examples indicate the tendency of water pentagons to form closed geometrical structures around amino acids (Fig. 6e,f) and nucleosides (Neidle, Berman, and Shieh 1980). Studies of supercooled water and amorphous ice have revealed the presence of cyclic and clathrate water pentamers (Nandi et al. 2017; Yokoyama, Kannami, and Kanno 2008), suggesting the possible delivery of water pentamers by ice-coated cosmic dust and meteorites to the atmospheres of Earth and habitable exoplanets. At the opposite extreme temperatures and pressures of hydrothermal ocean vents (*black smokers*) arising from planetary volcanic activity, the expelled water can be in the *supercritical* phase, where the structure is neither liquid nor vapor but instead isolated water nanoclusters (Sahle et al. 2013). Therefore, at both temperature and pressure extremes, water pentamers interacting with prebiotic molecules could nucleate additional water molecules - eleven in the Fig. 7 example - to form the more stable pentagonal dodecahedral cluster $(H_2O)_{21}H^+$, which could then provide via its THz vibrations (Fig. 2c,d) the eleven water molecules necessary to yield the RNA sequence of Fig. 7. In other words, water nanoclusters delivered by cosmic dust or hydrothermally to planet Earth and habitable exoplanets could have provided a catalytic pathway for the dehydration-condensation-reaction mechanism of RNA polymerization.

How a $(H_2O)_{21}H^+$ cluster could expel water molecules or OH groups when interacting with prebiotic molecules to promote RNA chain growth deserves further analysis. As pointed out in Section 2.2, this protonated water cluster readily takes up an extra electron into the LUMO (Fig. 2a) – a *hydrated electron* – as noted in Fig. 7. The proximity of the resulting electrically neutral $(H_2O)_{21}H$ cluster occupied LUMO "S" orbital to the lowest unoccupied, nearly degenerate cluster "$P_x,P_y,P_z$" orbitals (Fig. 2a,b) suggests the possible coupling between the hydrated electron and the pentagonal dodecahedral cluster THz-frequency "squashing" and "twisting" modes shown in Fig. 8a via the *pseudo* or *dynamic Jahn-Teller (JT) effect* (Bersuker and Polinger 1989).

JT coupling in $(H_2O)_{21}H$ leads to a prescribed symmetry breaking of the pentagonal dodecahedron along the THz-frequency vibrational mode coordinates $Q_s$, lowering the cluster potential energy from A to the equivalent minima A' shown in Fig. 8b. Because of the large JT-induced vibrational displacements (large $Q_s$) of water-cluster surface oxygen atoms, the energy barrier for expulsion of water oxygen or OH radicals and their oxidative addition to reactive nucleotides is lowered from $E_{barrier}$ to $E'_{barrier}$ (Fig. 8b).

**7.2. RNA protocells**

Laboratory investigations of the possible origins of life on Earth have successfully created self-assembling model protocell membranes, the simplest of which are fatty-acid vesicles capable of containing at least short segments of RNA (Meierhenrich et al. 2010; Dworkin et al. 2001; Oberholzer et al. 1995; Chang, Huang, and Hung 2000; Hanczyc and Szostak 2004). Fatty acids are *amphiphilic* molecules, which means that polar and nonpolar functional groups are present in the same molecule. Fatty acids are commonly found in experiments simulating the prebiotic "soup" arising from Earth's early hydrothermal conditions (Milshteyn et al. 2018) and the arrival of extraterrestrial material to the early Earth (Kvenvolden 1970). Quantum-chemical calculations by the SCF-Xα-Scattered-Wave density-functional method (Slater and Johnson 1972, 1974) find that the polar (*hydrophilic*) end of the naturally occurring fatty acid *glycerol monolaurate* (*"GML/monolaurin"*) $C_{15}H_{30}O_4$ (Milshteyn et al. 2018) attracts water molecules, forming a stable water-nanocluster-GML molecule, as illustrated in Fig. 9a. These calculations reveal: (1) the polar end of the fatty acid donates an electron into the water-cluster LUMO, as shown by the computed molecular-orbital wavefunction $\Psi$ in Fig. 9a; (2) the 1.8 THz vibrational mode of the water nanocluster described above resonates with the fatty-acid carbon-chain motions, as represented by the vectors in Fig. 9b. This quantum-mechanical coupling of the fatty acid to water nanoclusters can promote their chemical reactivity with the prebiotic organic molecules and phosphates leading to RNA polymerization by the steps shown in Fig. 7. Once such couplings occurred naturally during Earth's early hydrothermal conditions and the arrival of extraterrestrial material to the early Earth, the water-nanocluster-fatty-acid molecules tended to aggregate around growing RNA segments, as shown in Fig. 9c. The preferred orientation of these molecules around the RNA segment will be recognized as a primitive *reverse micelle*, the simplest self-assembling fatty-acid vesicle that has been demonstrated to be capable of containing at least short segments of RNA (Chang, Huang, and Hung 2000). The laboratory-controlled synthesis of self-assembling water-nanocluster reverse micelles present in *water-in-oil nanoemulsions* has also been demonstrated (Johnson 1998; Daviss 1999).

This scenario, albeit rudimentary, provides at least one possible pathway to cellular life's beginnings on Earth and habitable exoplanets, including the earliest RNA viruses (Moelling and Broecker 2019), while suggesting contemporary applications to biomedicine, such as pharmaceuticals (Authelin et al. 2014) and *RNA-interference antiviral drugs* (Wu and Chan 2006; Setten, Rossi, and Han 2019). RNA interference induced by small interfering RNA segments like the micellular one shown in Fig. 9c can inhibit the expression of viral antigens and so provides a novel approach to the therapy of pathogenic coronaviruses such as COVID-19.

# 8. Conclusions

## 8.1 Why pentagonal water clusters?

In the proposed roles of water nanoclusters in cosmology and astrobiology, emphasis has been placed on clusters consisting of pentagonal cyclic rings of water molecules. This is justified because a pentagonal geometry leads to the magic-number water nanoclusters most frequently observed experimentally and discussed in Section 2. The higher stability of these clusters is explained simply by the fact that the water molecule bond angle is roughly equal to a regular pentagon angle. Thus, the water molecule hydrogen bonds are only slightly deformed. That said, this scenario does not completely rule out the possible ejection of water nanoclusters of other topologies from cosmic dust according to Eq 1. For example, the calculated cut-off THz vibrational frequency of an icosahedral "water buckyball" is only slightly greater than that of a pentagonal dodecahedral one, and deforming the latter dodecahedron changes that frequency only marginally.

## 8.2 Can classical physics explain water nanoclusters?

In Section 5.1 it was pointed out that pentagonal dodecahedral water nanoclusters of the types shown in Figs. 2 and 3 might be viewed as spherical shells of the type originally proposed by Layzer and Hively (1973) to be a primordial source of the CMB radiation according to classical electromagnetic theory. Likewise, viewing such nanoclusters as tiny spheres could allow one, in principle, to apply the theoretical approach of Gerardy and Ausloos (1983,1984) to model the infrared absorption spectrum of water-nanocluster arrays from solutions of Maxwell's equations. However, instead of the infrared, here we are focused on the unique THz spectra of such nanoclusters due to the "surface" vibrational modes of the clusters' water-molecule shells.

Since the classical frequency of a thin vibrating spherical shell varies inversely with the radius of shell, it is therefore no surprise that the THz cut-off frequency of an approximately spherical water nanocluster decreases with increasing cluster size or with increasing number of water molecules in Eq. 1, as suggested by Figs 2 and 3. Likewise, since the classical frequency (lowest harmonic) of a vibrating string is inversely proportional to its length, "strings" of water nanoclusters like the one shown in Fig. 4 will tend to have cut-off frequencies decreasing with string length. The numerical values of these frequencies cannot be determined classically, but they can be computed quantum mechanically by the SCF-Xα-Scattered-Wave density-functional method (Slater and Johnson 1972, 1974), as described in Section 2.2.

## 8.3 The key frequency

My quantum-chemical findings that 1.8 THz water-nanocluster vibrational modes are key to their role in coupling with and activating prebiotic molecules is especially interesting because that frequency is very close to the cut-off frequency $\nu_c \cong 1.7$ THz which determines, according to Section 4 and Eq. 4, the present dark energy density due to vacuum fluctuations, and that is consistent with the measured cosmological constant at the present time in the expansion of the universe. This result is also consistent with water nanoclusters containing the "magic-numbers" of n = 21 and 20 water molecules (Figs. 2d and 3a, respectively) to be dominant ones produced by ice-coated cosmic dust according to Eq. 1, because with increasing cluster size (Figs. 3b,c and 4), the cut-off frequency $\nu_c$ and, according to Eq. 4, the dark energy density will be smaller than presently measured. Applying the cyclic cosmology in Section 5.4 and the astrobiology in Section 6, I conclude that as the universe expands further and larger cosmic water nanoclusters become more dominant from ice-coated cosmic dust, those larger clusters will be less favorable to interact with prebiotic molecules, suggesting that life will become less probable over astronomical time. In other words, we are likely living at the ideal time in the expanding universe for life - as we know it - to exist, and water nanoclusters of the types shown in Figs. 2d and 3a created on cosmic dust could possibly be "seeds of life" that catalyze the biomolecules necessary for life.

## 8.4 Evidence for quintessential birefringent cosmic water nanoclusters

As discussed in Section 2, there is strong experimental evidence for the existence of water nanoclusters and protonated water-nanocluster cluster ions in Earth's atmosphere (Aplin and McPheat 2005), produced in laboratory vacuum chambers (Johnson et al. 2008) (Fig. 1), and generated from amorphous ice by energetic ion bombardment (Martinez et al. 2019). The protonated $(H_2O)_{21}H^+$ or $H_3O^+(H_2O)_{20}$ cluster (Fig 2) is exceptionally stable, even at high temperatures and irradiation, and has been identified experimentally from its infrared spectrum [Miyazaki et al. 2004; Shin et al. 2004). Because this species can be viewed as a hydronium ion ($H_3O^+$) caged by a pentagonal dodecahedron of twenty water molecules, the report of widespread hydronium in the galactic medium through its inversion spectrum (Lis et al. 2014) is key to the challenge of identifying the presence of $H_3O^+(H_2O)_{20}$ because both spectra fall into the same THz region.

Recent evidence for *cosmic birefringence* extracted from the 2018 Planck CMB polarization data (Minami and Komatsu 2020) may possibly support the proposed *quintessence* scenario due to the birefringence property of cosmic water nanoclusters. Fig. 10 depicts $H_3O^+(H_2O)_{20}$ in the terahertz-induced vibrational state that has been shown in Sections 8.3 and 7.2 to be key to its "finely-tuned" coupling with prebiotic molecules and to RNA protocell creation. The transient distortion or symmetry breaking of the water dodecahedron via the dynamical Jahn-Teller effect (Bersuker and Polinger 1989) leads to the indicated *anisotropic dipole moments* computed by the SCF-Xα cluster molecular-orbital density-functional method (Slater and Johnson 1972, 1974). Anisotropic dipole moments can lead, in principle, to the birefringence of water nanoclusters – like that observed in the THz-induced birefringence of liquid water (Zhao *et al.* 2020) and other polar liquids (Sajadi *et al.* 2017).

## 8.5 Final conclusions

Although this paper is "out of the box" of generally popular inflationary cosmology and multiverse theory (Guth 1981, 2007; Linde 2008), it is not the only example. Other "cyclic-universe" theories have been widely promoted (Ijjas, Steinhardt, and Loeb 2017; Ijjas and Steinhardt 2019; Penrose 2006), while critiquing the multiverse scenario (Steinhardt 2011). This author has offered a unified *interdisciplinary* approach to cyclic cosmology and the origin of life in the universe based on quantum astrochemistry, while still attempting to include complementary relevant astrophysics facts. Certainly, not all problems in cosmology are solved by cosmic water nanoclusters, but perhaps some. The proposal that cosmic water nanoclusters may constitute a form of invisible baryonic dark matter does not rule out nonbaryonic dark matter, such as WIMPS and AXIONS, although observational evidence for their existence is lacking (Haynes 2018). Regarding dark energy, it is further proposed that cosmic water nanoclusters constitute a *quintessence scalar field* (Steinhardt 2003) that permeates the vacuum of space and largely cancels via Eq. 3 the otherwise infinite vacuum energy density, Eq. 2, at this point in time. This scenario is consistent with the reported "web" of dark matter permeating the cosmos (Heymans et al. 2012). The report of a neutral hydrogen gas bridge connecting the Andromeda (M31) and Triangulum (M33) galaxies (Lockman, Free, and Shields 2012) suggests the possibility that water nanoclusters, albeit at lower density than pure hydrogen, might similarly be dispersed as intergalactic gas constituting Rydberg dark matter. The striking consistency of the THz cut-off vibrational frequencies of water nanoclusters with the zero-point vacuum energy THz cut-off frequency that produces a dark-energy density in agreement with cosmological data is possibly only coincidental. Nevertheless, a common origin of dark matter and dark energy outside the realm of conventional elementary-particle physics, which has yet to identify conclusively the origins of either dark matter or dark energy, is a tempting idea. Indeed, one might also conceptually view water nanoclusters in the vacuum of space as baryonic nanoparticles that break the real-space symmetry of the otherwise isotropic vacuum due to their physical presence. Surprisingly, their masses are within the range of those estimated for WIMPS. Again, there are no other identified baryonic substances in our universe, including hydrogen, water monomers, and organic molecules that exhibit all these characteristics, while possessing the Rydberg-excited electronic states of low-density condensed matter that qualifies also as dark matter. Fullerene buckyballs in planetary nebulae (Cami et al. 2010) are also ruled out as candidates, even though Rydberg states have been observed in the $C_{60}$ molecule (Boyle et al. 2001) because their vibrational frequencies lie beyond the required THz range. Planets, and moons, as well as water vapor in solar atmospheres (Bergin and van Dishoeck 2012), nebulae (Glanz 1998), and distant quasars (Bradford et al. 2011) are widely present throughout the cosmos, and therefore should be included as possible sources of cosmic water nanoclusters. In fact, water clusters have been detected recently in the hydrothermal plume of Enceladus – a moon of Saturn (Coates et al. 2013). Finally, from the quantum chemistry of cosmic water nanoclusters interacting with prebiotic organic molecules, amino acids and RNA protocells on early Earth and habitable exoplanets, this scenario is consistent with the *anthropic principle* that our universe is a connected biosystem and has those properties which allow life, as we know it - based on water, to develop at the present stage of its history.

**Acknowledgements**

I am grateful to Franziska Amacher for introducing me to the RNA world and for the continual support of Henry Johnson. I also thank the referees for their helpful suggestions.

**References**

Ade, P. A. R. et al. (2016) Planck 2015 results-xii. cosmological parameters. *A&A* 594, 1-28.
Akerib, D. S. et al. (2017) Results from a search for dark matter in the complete LUX exposure. *Phys. Rev. Lett.* 118, 021303-021311.
Aplin, K. L. and McPheat, R. A. (2005) Absorption of infra-red radiation by atmospheric molecular cluster-ions. *J. Atmos. Solar Terrest. Phys.* 67, 775-783.
Authelin, J. -R. et al. (2014) Water Clusters in amorphous pharmaceuticals. *J. Pharma. Sci.* 103, 2663-2672.
Badiei, S. and Homlid, L. (2002) Rydberg matter in space: low-density condensed dark matter. *Mon. Not. R. Astron. Soc.* 333, 360-364.
Bally, J. and Harrison, J. R. (1978) The electrically polarized universe. *ApJ* 220, 743-744.
Banandos, E. et al. (2019) A metal-poor damped Lyα system at redshift 6.4. *ApJ* 885, 59-74.
Beck, C. and Mackey, M. C. (2005) Could dark energy be measured in the lab? *Phys. Lett. B* 605, 295-300.
Beck, C. and Mackey, M. C. (2007) Measurability of vacuum fluctuations and dark energy. *Physica A* 379, 101-110.
Beck, C. and Mackey, M. C. (2006) Rebuttal to: Has dark energy really been discovered in the Lab? arXiv astro-ph/0603397, 1-5.
Bennett, C. L. et al. (2013) Nine-year Wilkinson microwave anisotropy probe (WMAP) observations: cosmological parameter results. *ApJS* 208, 1-25.
Bergin, E. A. and van Dishoeck, E. F. (2012) Water in star- and planet-forming regions. *Phil. Trans. R. Soc. A* 370, 2778-2802.
Bersuker, I. B. and Polinger, V. Z. (1989) *Vibronic Interactions in Molecules and Crystals*, Springer-Verlag: Berlin.
Bialy, S., Sternberg, A., and Loeb, A. (2015) Water formation during the epoch of first metal enrichment. *ApJ* 804, L29-L34.
Boyle, M. et al. (2001) Excitation of rydberg series in $C_{60}$. *Phys. Rev. Lett.* 83, 273401-273405.
Bradford, C. M. et al. (2011) The water vapor spectrum of APM 08279+5255: x-ray heating and infrared pumping over hundreds of parsecs. *ApJ* 741, L37-L43.
Brudermann, J., Lohbrandt, P., and Buck, U. (1998) Surface vibrations of large water clusters by He atom scattering. *Phys. Rev. Lett.* 80, 2821-2824.
Cafferty, B. J. and Hud, N. V. (2014) Abiotic synthesis of RNA in water: a common goal of prebiotic chemistry and bottom-up synthetic biology, *Current Opinions in Chemical Biology* 22, 146-157.
Cami, J. et al. (2010) Detection of C60 and C70 in a young planetary nebula. *Science* 329, 1180-1182.
Carlon, H. R. (1981) Infrared absorption by molecular clusters in water vapor. *J. Appl. Phys.* 52, 3111-3115.

Case, D. A., Huynh, B. H., and Karplus, M. (1979) Binding of oxygen and carbon monoxide to hemoglobin. An analysis of the ground and excited states. *JACS* 101, 4433-4453.
Chakraborty, K. et al. (2014) Possible features of galactic halo with electric field and observational constraints. *Gen.Relativ.Gravit.* 46, 1807-1820.
Chang, G. -G., Huang, T. -M., and Hung, H. -C. (2000) Reverse micelles as life-mimicking systems. *Proc. Nat. Sci. Counc. ROC(B)* 24, 89-100.
Chaplin, M. (2006) Do we underestimate the importance of water in cell biology? *Nature Rev. Molec. Cell Biol.* 7, 861-866.
Clowe, D. Gonzalez, A., and Markevich, A. (2004) Weak-lensing mass reconstruction of the interacting cluster 1E 0657-558: direct evidence for the existence of dark matter. *ApJ* 604, 596-604.
Coates, A. J. et al. (2013) Photoelectrons in the Enceladus Plume. *J. Geophys. Res. Space Phys*. 118, 5099-5108.
Costanzo, G. et al. (2009) Generation of long RNA chains in water. *J. Biol. Chem.* 284, 33206-33216.
Cotton, F. A., Norman, J. G., and Johnson, K. H. (1973) Biochemical importance of the binding of phosphate by arginyl groups. Model compounds containing methylguanidinium io*n*. *JACS* 95, 2367-2369.
Daviss, B. (1999) Just add water. *New Scientist*, 13 March.
Duley, W. W. (1996) Molecular clusters in interstellar clouds. *ApJ* 471, L57-L60.
Dulieu, F. et al. (2010) Experimental evidence for water formation on interstellar dust grains by hydrogen and oxygen atoms. *A&A* 512, A30.
Dworkin, J. P. et al. (2001) Self-assembling amphiphilic molecules: synthesis in simulated interstellar/precometary ices. *PNAS* 98, 815-819.
Gérardy, J. M. and Ausloos, M. (1983) Absorption spectrum of clusters of spheres from the general solution of Maxwell's equations. IV. Proximity, bulk, surface, and shadow effects (in binary clusters). *Phys Rev. B* 27, 6446-6463.
Gérardy, J. M. and Ausloos, M. (1984) Absorption spectrum of clusters of spheres from the general solution of Maxwell's equations. III. Heterogeneous spheres. *Phys. Rev. B* 30, 2167-2181.
Ghosh, J., Methikkalam, R. J., and Bhuin, R. G. (2019) Clathrate hydrates in the interstellar environment. *PNAS* 116, 1526-1531.
Glanz, J. (1998) A water generator in the Orion nebula. *Science* 280, 378-382.
Guth, A. H. (1981) Inflationary universe: a possible solution to the horizon and flatness problems. *Phys. Rev. D* 23, 347-356.
Guth, A. H. (2007) Eternal inflation and its implications. *J. Phys. A* 30, 6811-6826.
Hanczyc, M. M. and Szostak, J. W. (2004) Replicating vesicles as models of primitive cell growth and division. *Current Opinions in Chemical Biology* 8, 660-664.
Harker, H. A. et al. (2005) Water pentamer: characterization of the torsional-puckering manifold by terahertz VRT spectroscopy. *J. Phys. Chem. A* 109, 6483-6497.
Haynes, K. (2018) What is dark matter? Even the best theories are crumbling. *Discover Magazine*, September 21.
Herzberg, G. (1987) Rydberg molecules. *Ann. Rev. Phys. Chem.* 38, 27-56.
Heymans, C. et al. (2012) CFHTLenS: the Canada-France-Hawaii telescope lensing survey. *Mon. Not. R. Astron. Soc.* 427, 146-166.
Holmlid, L (2008) Vibrational transitions in Rydberg matter clusters from stimulated Raman and Rabi-flopping phase delay in the infrared. *J. Raman Spectroscopy* 39, 1364-1374.
Iglesias-Groth, S. et al. (2010) A search for interstellar anthracene towards the Perseus anomalous microwave emission region*. Mon. Not. R. Astron. Soc.* 407, 2157-2165.
Ijjas, A. and Steinhardt, P. J. (2019) A new kind of cyclic universe. *Phys. Lett. B* 795, 666-672.
Ijjas, A., Steinhardt, P. J., and Loeb, A. (2017) Pop goes the universe. *Sci. Am.* 316, 32-39.
Jedamzik, K. and Pogosian, L. (2020) Relieving the Hubble tension with primordial magnetic fields. arXiv:2004.09487v2 [astro-ph.CO] Apr. 28.
Jetzer, P. and Straumann, N. (2005) Has dark energy really been discovered in the Lab? *Phys. Lett. B* 606, 77-79.
Jetzer, P. and Straumann, N. (2006) Josephson junctions and dark energy. *Phys. Lett. B* 639, 57-59.

Johnson, K. (2021) Cosmology, astrobiology, and the RNA world: just add quintessential water. *International Journal of Astrobiology* 20, 111-124. arXiv:2012.1 2079v2.

Johnson, K. (2012) Terahertz vibrational properties of water nanoclusters relevant to biology. *J. Biol. Phys.* 38, 85-95.

Johnson, K. H. et al. (2008) Water vapor: an extraordinary terahertz wave source under optical excitation. *Phys. Lett. A* 371, 6037-6040.

Johnson, K. H. (1998) Water clusters and uses therefor. *U.S. Patent No. 5,800,576*.

Jordan, K. D. (2004) A fresh look at electron hydration. *Science* 306, 618-619.

Joyce, G. F. and Orgel, L. E. (1993) Prospects for understanding the origin of the RNA world. In: Gesteland, R. F., Atkins, J. F., editors. *The RNA World*, Cold Spring Harbor Laboratory Press: Cold Spring Harbor, NY. pp. 1-22.

Kvenvolden, K. A. et al. (1970) Amino acids in the Murchison meteorite. *Nature* 228, 923-926.

Lacy, J. H. et al. (1991) Discovery of interstellar methane – observations of gaseous and solid $CH_4$ absorption toward young stars in molecular clouds. *ApJ* 376, 556-560.

Laporte, N. et al. (2017) Dust in the reionization era: ALMA observations of a $z = 8.38$ gravitationally lensed galaxy. *ApJ* 837, L21-L27.

Layzer, D. and Hively, R. (1973) Origin of the microwave background. *ApJ* 179, 361-370.

Lehnert, M. D. et al. (2010) Spectroscopic confirmation of a galaxy at redshift $z = 8.6$. *Nature* 467, 940-942.

Leonhardt, U. 2020 The case for a Casimir cosmology. *Phil. Trans. R. Soc.* A 378, 2019.0229

Linde, A. D. (2008) Inflationary cosmology. In: *Inflationary Cosmology*, Springer: Berlin.

Linder, E. V. and Jenkins, A. (2003) Cosmic structure growth and dark energy. *Mon. Not. Astron. Soc.* 346, 573-583.

Lis, D. C. et al. (2014) Widespread rotationally hot hydronium ion in the galactic interstellar medium. *ApJ* 785, 135-144.

Lockman, F. J., Free, N. L., and Shields, J. C. (2012) The neutral hydrogen bridge between M31 and M33. *Astron. J.* 144, 52-59.

Mann, I. (2001) Spacecraft charging technology. In: Harris, R. A., editor. *Proceedings of the Seventh International Conference, April 23-27.,* European Space Agency, ESA SP-476. pp.629-639.

Martinez, R. et al. (2019) Production of hydronium ion $(H3O)+$ and protonated water clusters $(H2O)nH+$ after energetic ion bombardment of water ice in astrophysical environments. *J. Phys. Chem. A* 123, 8001-8008.

Matsuura, M. et al. (2019) SOFIA mid-infrared observations of supernova 1987A in 2016 – forward shocks and possible dust re-formation in the post-shocked region. *Mon. Not. R. Astron. Soc.* 482, 1715-1723.

Meierhenrich, U. J. et al. (2010) On the origin of primitive cells: from nutrient intake to elongation of encapsulated nucleotides. *Angew.Chem.Int.Ed.* 49, 3738-3750.

Milshteyn, D. et al. (2018) Amphiphilic compounds assemble into membranous vesicles in hydrothermal hot spring water but not in seawater. *Life* 8, 11-26.

Minami, Y. and Komatsu, E. (2020) New extraction of the cosmic birefringence from the Planck 2018 polarization data. *Phys. Rev. Lett.* 125, 221301-1-221301-6.

Miyazaki, M. et al. (2004) Infrared spectroscopic evidence for protonated water clusters forming nanoscale cages. *Science* 304, 1134-1137.

Moelling, K. and Broecker, F. (2019) Viruses and evolution – viruses first? A personal perspective. *Frontiers in Microbiology* 10, 1-13.

Munoz, J. B. and Loeb, A. (2018) A small amount of mini-charged dark matter could cool the baryons in the early universe, *Nature* 557, 684-686.

Nandi, P. K. et al. (2017) Ice-amorphization of supercooled water nanodroplets in no man's land. *ACS Earth and Space Chemistry* 1, 187-186.

Neidle, S., Berman, H., and Shieh, H. S. (1980) Highly structured water network in crystals of deoxydinucleoside-drug complex. *Nature* 288, 129-133.

Oberholzer, T. et al. (1995) Enzymatic RNA replication in self-reproducing vesicles: an approach to a minimal cell. *Biochem.Biophys.Res.Commun*. 207, 250-257.

Oesch, P. A. et al. (2016) A remarkably luminous galaxy at $z = 11.1$ measured with Hubble space telescope Grism spectroscopy. *ApJ* 819, 129-140.

Ouellet, et al. (2019) First results from ABRACADABRA-10 cm: a search for sub-µeV axion dark matter. *Phys. Rev. Lett.* 122, 121802-121809.

Penrose, R. (2006) Before the big bang: an outrageous new perspective and its implications for particle physics. In: *Proceedings of EPAC 2006, Edinburgh, Scotland*. pp. 2759-2762.

Potapov, A., Jager, C., and Henning, T. (2020) Ice coverage of dust grains in cold astrophysical environments. *Phys. Rev. Lett.* 124, 221103-1-221103-7.

Ratra, P. and Peebles, L. (1988) Cosmological consequences of a rolling homogeneous scalar field. *Phys Rev. D* 37, 3496-3427.

Riess, A. G. et al. (1998) Observational evidence from supernovae for an accelerating universe and a cosmological constant. *Astron. J.* 116, 1009-1038.

Risaliti, G. and Lusso, E. (2019) Cosmological constraints from the Hubble diagram at high redshifts. *Nature Astron.* 3, 272-277.

Sahle, C. J. et al. (2013) Microscopic structure of water at elevated pressures and temperatures. *PNAS* 110, 6301-6306.

Sajadi, M., Wolf, M. and Kampfrath, T. (2017) Transient birefringence of liquids induced by terahertz electric-field torque on permanent molecular dipoles. *Nature Communications* 8,14963-14970. arXiv:1610.04024.

Setten, R. L., Rossi, J. J., and Han, S. P. (2019) The current state and future directions of RNAi-based therapeutics. *Nature Reviews Drug Discovery* 18, 421-446.

Shin, J. W. et al. (2004) Infrared signature of structures associated with $H^+(H_2O)n$ (n = 6 to 27). *Science* 304, 1137-1140.

Slater, J. C. and Johnson, K. H. (1972) Self-consistent-field Xα cluster method for polyatomic molecules and solids. *Phys. Rev. B* 5, 844-853.

Slater, J. C. and Johnson, K. H. (1974) Quantum chemistry and catalysis. *Physics Today* 27, 34-41.

Sola, J. (2014) Vacuum energy and cosmological evolution. *AIP Conference Proceedings* 1606, 19-37.

Souza, R. D., Impens, F., and Neto, P. A. M. (2018) Microscopic dynamical Casimir effect. *Phys. Rev. A* 97, 032514-032523.

Steinhardt, P. J. (2003) A quintessential introduction to dark energy. *Phil. Trans. R. Soc. Lond. A* 361, 2497-2513.

Steinhardt, P. J. (2011) Inflation theory debate: is the theory at the heart of modern cosmology deeply flawed? *Sci. Am.* 304N4, 18-25.

Teeter, M. M. (1984) Water Structure of a hydrophobic protein at atomic resolution: pentagon rings of water molecules in crystals of crambin. *Proc. Natl. Acad. Sci.* 81, 6014-6018.

Totani, T. (2020) Emergence of life in an inflationary universe. *Scientific Reports* 10, 1671-1678.

Weinberg, S. (1987) Anthropic bound on the cosmological constant. *Phys. Rev. Lett.* 59, 2607-2610.

Weinberg, S. (1989) The cosmological constant problem. *Rev. Mod. Phys.* 61, 1-23.

Weinberg, S. (2008) *Cosmology,* Oxford University Press: New York. pp. 185-200.

Wright, E. L. (1982) Thermalization of starlight by elongated grains – could the microwave background have been produced by stars. *ApJ* 255, 401-407.

Wu, C. -J. and Chan, Y. -L. (2006) Antiviral applications of RNAi for coronavirus. *Expert Opin. Investig. Drugs* 15, 89-96.

Yang, C. Y., Johnson, K. H., Holm, R. H., and Norman, J. G. (1975) Theoretical model for the 4-Fe active sites in oxidized ferredoxin and reduced high-potential proteins. Electronic structure of the analog $[Fe_4S^*_4(SCH_3)_4]^{2-}$. *JACS* 97, 6596-6598.

Yokoyama, H., Kannami, M., and Kanno, H. (2008) Intermediate range O-O correlations in supercooled water. *Chem. Phys. Lett.* 463, 99-102.

Zeldovich, Ya. B. and Krasinski, A. (1968) The cosmological constant and the theory of elementary particles. *Sov. Phys. Usp.* 11 381-393.

Zhao, H., Tan Y., Zhang L., Zhang, R., Shalaby, M., Zhang, C., Zhao, Y. and Zhang, X.-C. (2020) Ultrafast hydrogen bond dynamics of liquid water revealed by terahertz-induced transient birefringence. *Light: Science & Applications* 9:136, 1-10.

Zheng, W. et al. (2012) A magnified young galaxy from about 500 million years after the big bang. *Nature* 489, 406-408.

# Figure Captions

**Fig. 1.** Frequency dependence for a range of pressures of the THz wave generation amplitudes (in arbitrary units) from water nanoclusters produced from water vapor, described in publication (Johnson et al. 2008). For pressure less than 20 Torr, the data were taken from a cell containing water vapor, while at higher pressure (>20 Torr), the data were taken from the pulsed nozzle.

**Fig. 2.** Molecular orbital energy levels, wavefunctions, and vibrational modes of the $(H_2O)_{21}H^+$ cluster calculated by the SCF-Xα-Scattered-Wave density-functional method (Slater and Johnson 1972, 1974). **a.** Cluster molecular-orbital energy levels. The HOMO-LUMO energy gap is approximately 3 eV. **b.** Wavefunctions of the lowest unoccupied cluster molecular orbitals. **c.** THz vibrational spectrum. **d.** Lowest-frequency THz vibrational mode and relative vibrational amplitudes. The vectors show the directions and relative amplitudes for the oscillation of the hydronium ($H_3O^+$) oxygen ion coupled to the O–O–O "bending" motions of the cluster "surface" oxygen ions.

**Fig. 3. a.** THz vibrational spectrum of a pentagonal dodecahedral $(H_2O)_{20}$ cluster calculated by the SCF-Xα-Scattered-Wave density-functional method (Slater and Johnson 1972, 1974). **b.** Lowest-frequency THz vibrational mode and relative vibrational amplitudes. **c.** Calculated THz vibrational spectrum of a stable array of three dodecahedral water clusters. **d**. Lowest-frequency THz vibrational mode. **e.** THz vibrational spectrum of a stable array of five dodecahedral water clusters. **f.** Lowest-frequency THz vibrational mode.

**Fig. 4. a.** Stable linear array of five pentagonal dodecahedral water nanoclusters. **b.** THz vibrational spectrum calculated by the SCF-Xα-Scattered-Wave density-functional method (Slater and Johnson 1972, 1974). **c.** Lowest-frequency THz vibrational mode and relative vibrational amplitudes. Note the lower cut-off THz frequency as compared with Fig. 3e,f.

**Fig. 5.** Microscopic dynamical Casimir absorption of zero-point-energy vacuum fluctuation "virtual" photons between THz vibrational states of a cosmic water nanocluster. The fact that the nanocluster "mechanical" THz vibrational frequencies are smaller than the optical transition frequencies leads to the emission of "real" photons (Souza, Impens, and Neto 2018).

**Fig. 6. a.** Pentagonal dodecahedral $CH_4(H_2O)_{20}$ methane clathrate. **b.** 1.8 THz vibrational mode. **c.** $(H_2O)_{20}$ cluster interacting with anthracene. **d.** The vibrational coupling of the $(H_2O)_{20}$ 1.8 THz vibrational mode to a THZ "bending" mode of anthracene. **e.** Hemispherical pentagonal dodecahedral water nanocluster "clathrating" a valine amino acid residue. **f.** Vibrational coupling of the water clathrate 1.8 THz vibrational mode to the valine amino acid "bending" mode.

**Fig. 7.** Dehydration-condensation exchange of eleven water molecules between prebiotic nucleosides and phosphates to form an RNA sequence of four polymerized nucleobases, *guanine*, *adenine*, *uracil*, and *cytosine* via a water nanocluster catalytic pathway for the water molecules This may explain how the first self-replicating RNA polymers were created chemically from a pool of prebiotic organic molecules and thereby clarify the "RNA-World" origin of life – preceding DNA- and protein-based life - on planet Earth and exoplanets in the habitable zones of developed solar systems wherever water is present.

**Fig. 8. a.** "Squashing" and "twisting" vibrational modes of a pentagonal dodecahedron, where $H_g$ and $H_u$ designate the key irreducible representations of the icosahedral point group corresponding to these modes. **b.** Schematic representation of the double potential energy wells for a Jahn-Teller distorted water pentagonal dodecahedral water nanocluster and the resulting reduction of the energy barrier for the catalytic reaction of the cluster along the reaction path defined by the normal mode coordinates $Q_s$.

**Fig. 9. a.** Water nanocluster attached to the polar end of the naturally occurring fatty acid *glycerol monolaurate* (*"GML/monolaurin"*) $C_{15}H_{30}O_4$, which donates an electron to the water-cluster LUMO represented by the computed molecular-orbital wavefunction $\Psi$. **b.** The 1.8 THz vibrational mode of the combined water-nanocluster-GML molecule. Such quantum-mechanical coupling of fatty acids to water nanoclusters can promote their chemical reactivity with the prebiotic organic molecules and phosphates, leading to RNA polymerization by the steps shown in Fig. 7. **c.** The aggregation of water-nanocluster-GML molecules around an RNA segment, forming a primitive reverse micelle.

**Fig. 10. a.** 1.8 THz vibrational mode of the cosmic water nanocluster $H_3O^+(H_2O)_{20}$ holding a *hydrated electron* - key to 'fine tuning' prebiotic molecules for life and nearly equal in frequency to the value $\nu_c \cong 1.7$ THz that determines, according to Section 4, the present cosmic dark energy density. The vectors represent the vibrational directions and amplitudes of the "clathrated" hydronium $H_3O^+$ oxygen ion coupled with the O–O–O "bending" motions of the cluster "surface" oxygen ions. The anisotropic dipole moments (in Debyes) along the nanocluster axes are indicated and are precursors to water nanocluster birefringence analogous to the THz-induced birefringence of liquid water (Zhao *et al.* 2020) and other polar liquids (Sajadi *et al.* 2017). **b.** Wavefunction $\Psi$ of the attached electron in this vibrational state - key to the "finely-tuned" coupling with prebiotic molecules. The cooperative birefringence of quantum-entangled water nanoclusters ejected to interstellar space from water-ice-coated cosmic dust at redshift $z \cong 10$ may contribute to the CMB and possibly explain, at least qualitatively, the cosmic birefringence observations of Minami and Komatsu (2020).

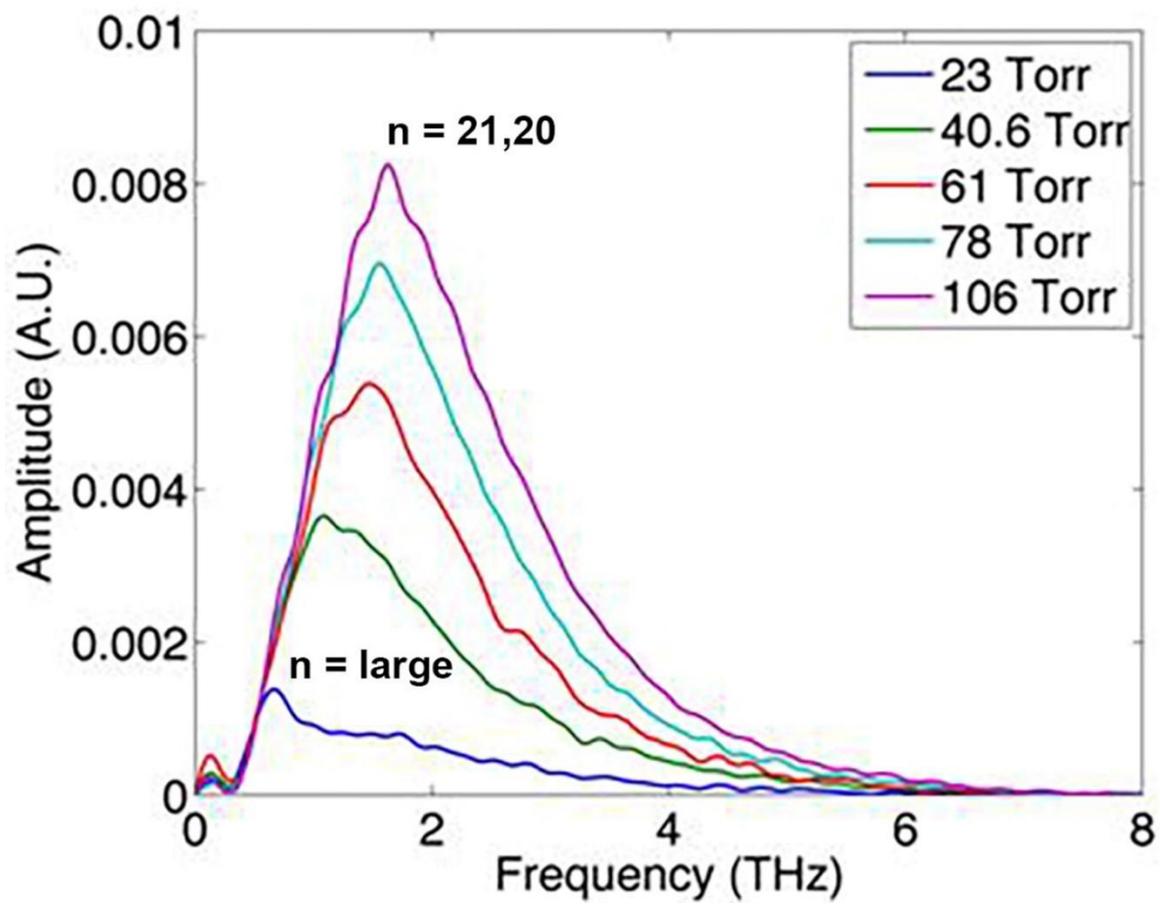

**Fig. 1**

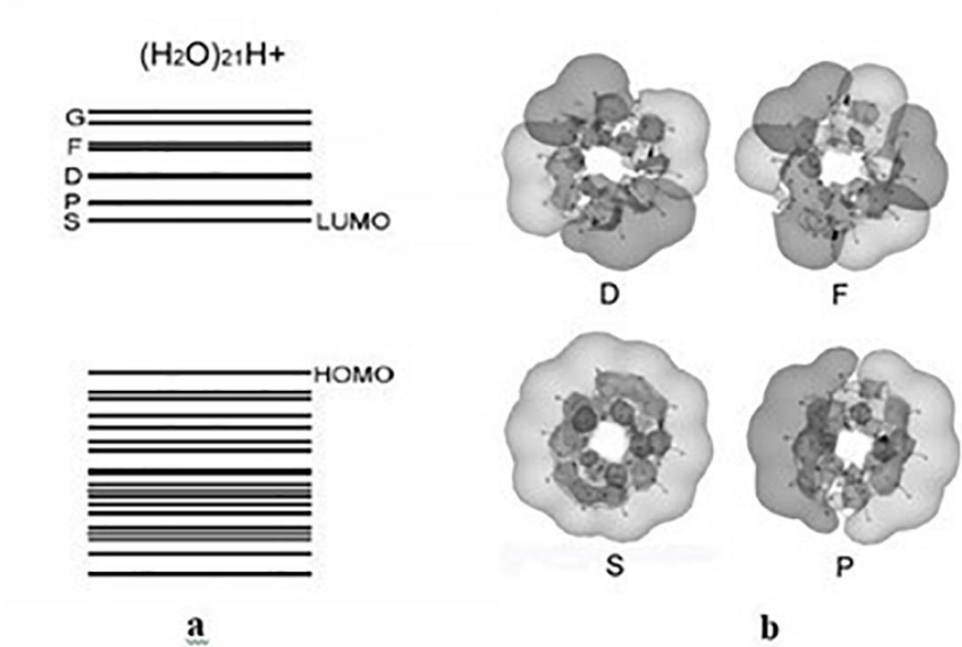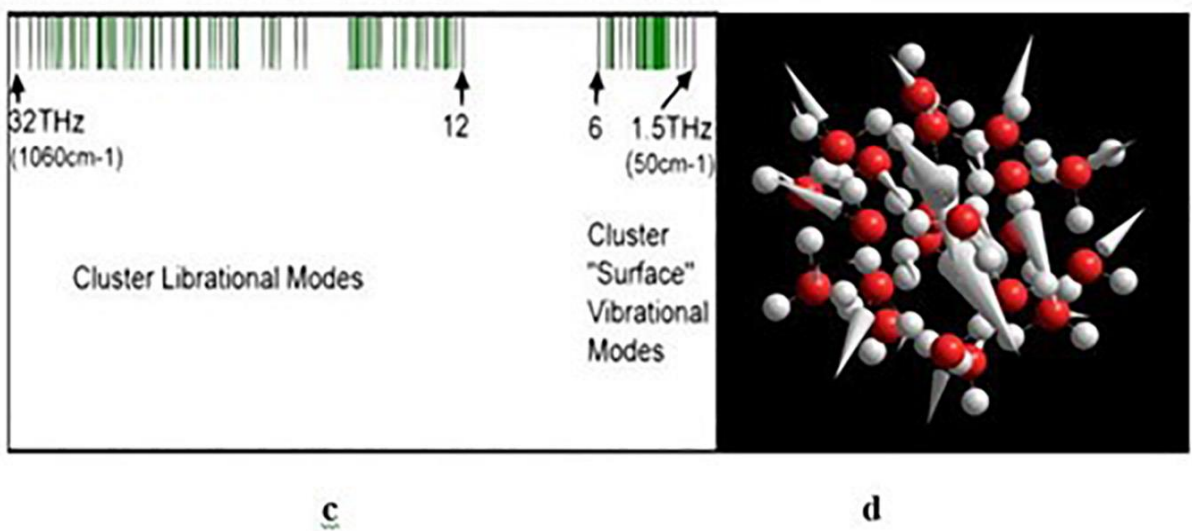

Fig. 2

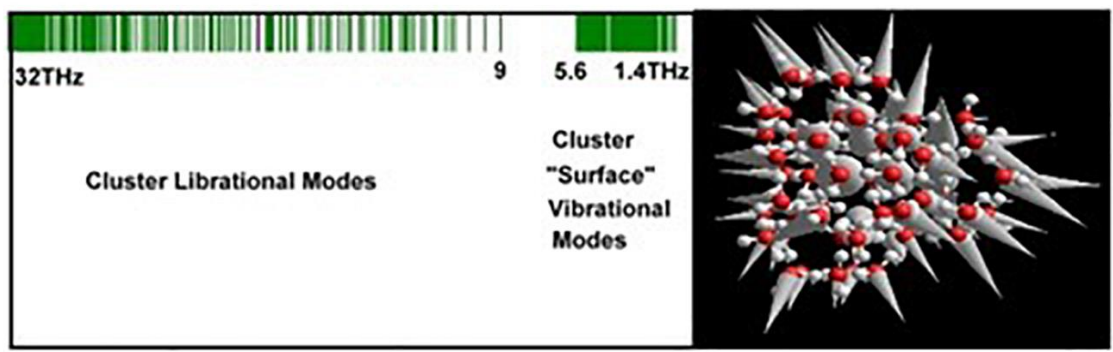

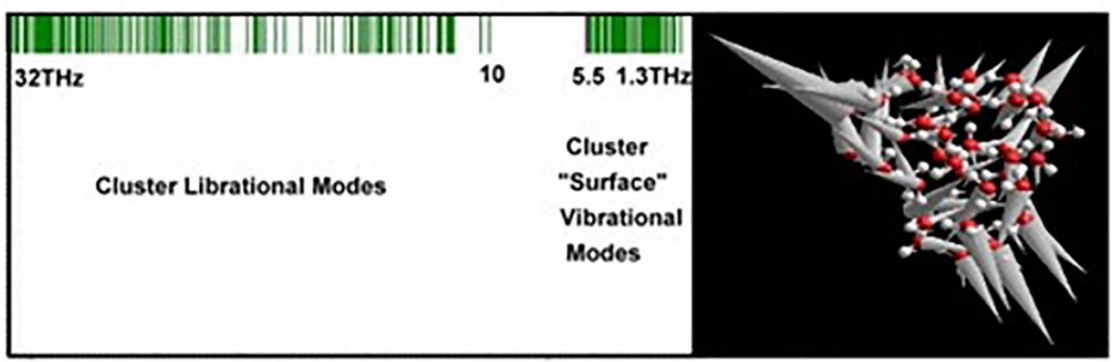

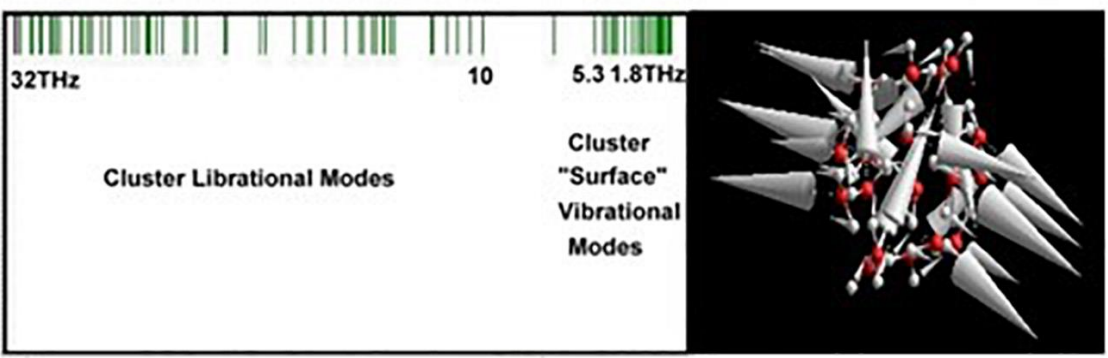

**Fig. 3**

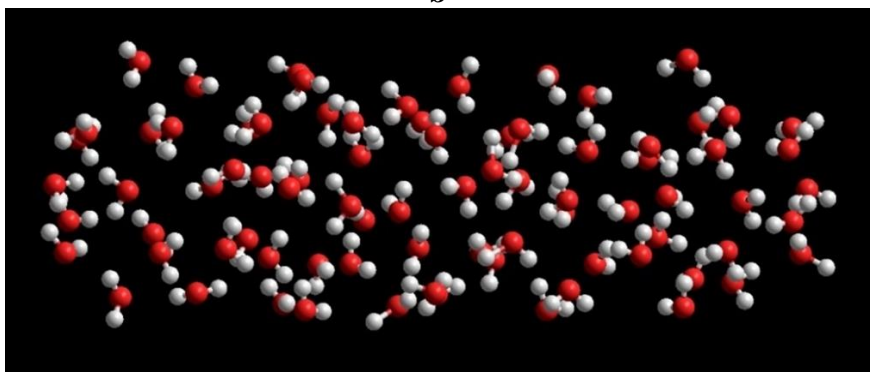

c

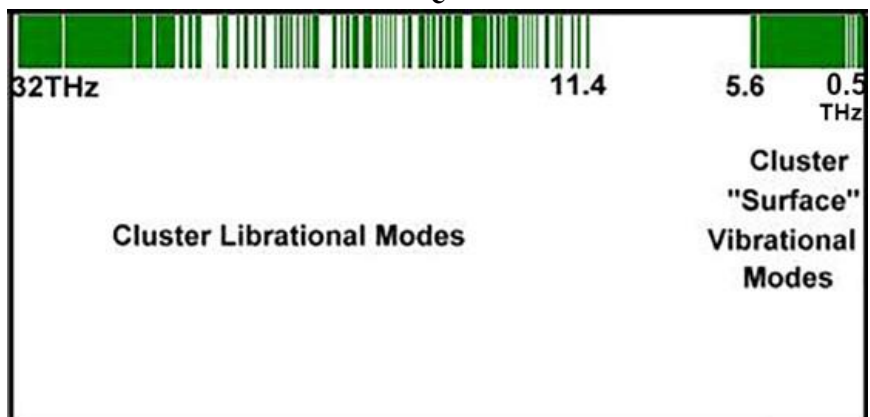

b

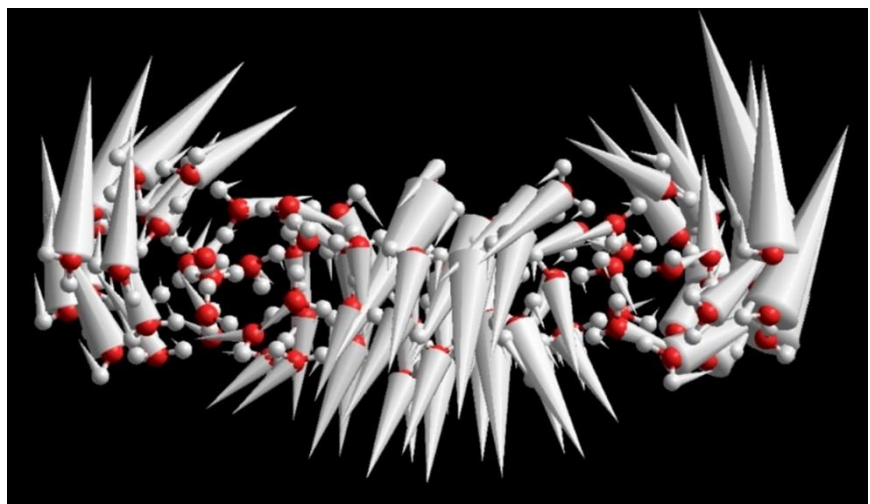

a

**Fig. 4**

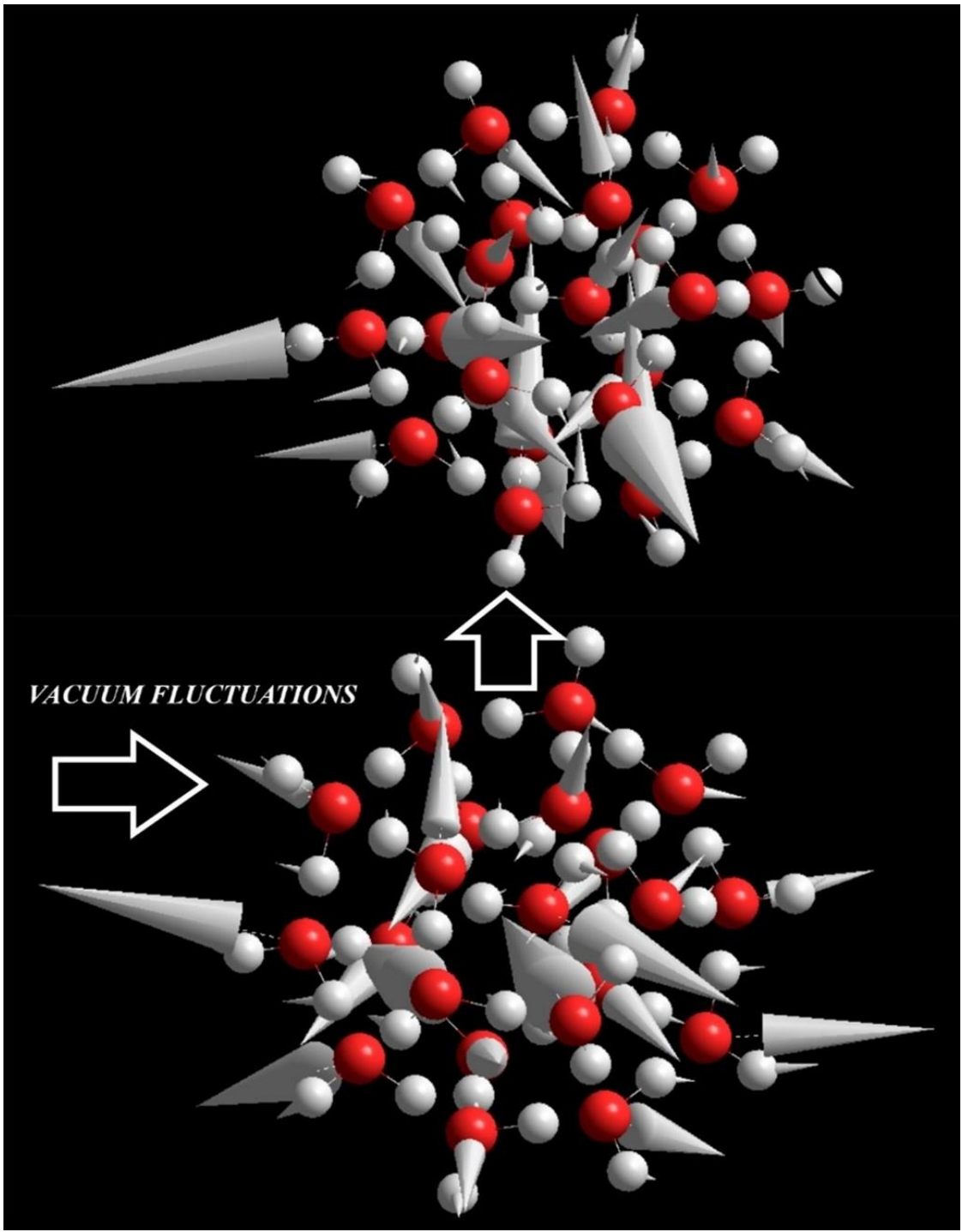

**Fig. 5**

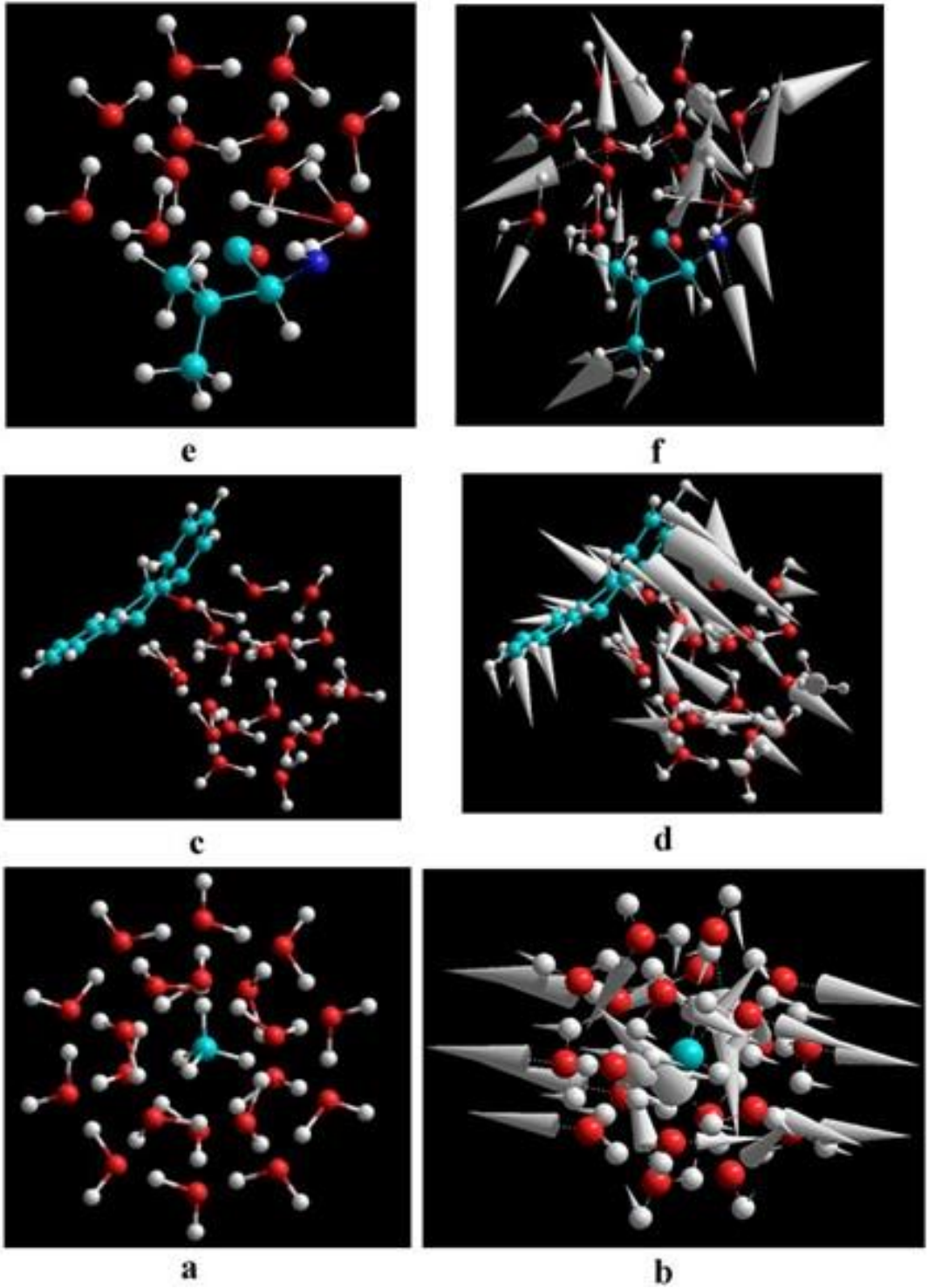

**Fig. 6**

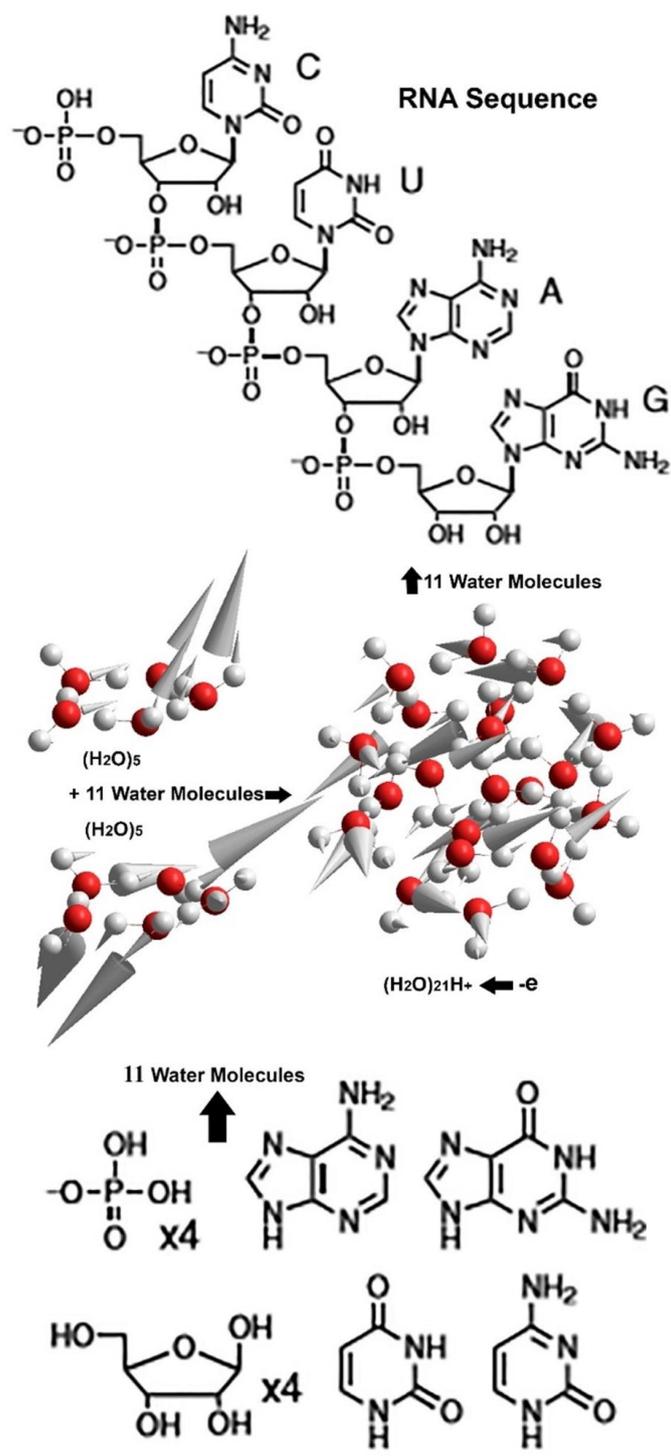

**Fig. 7**

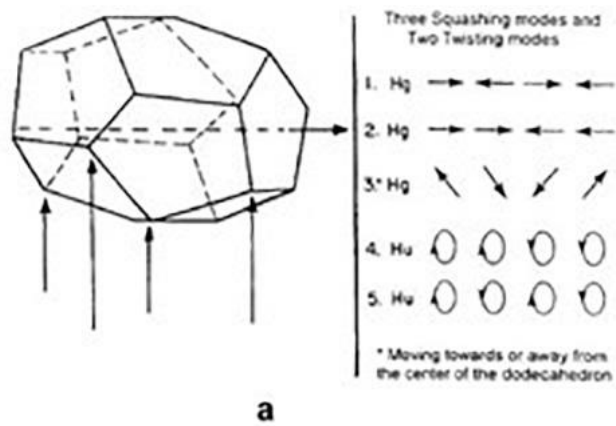

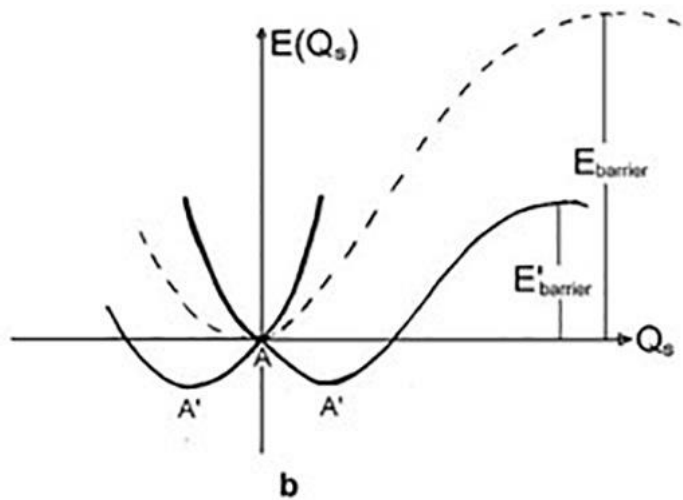

**Fig. 8**

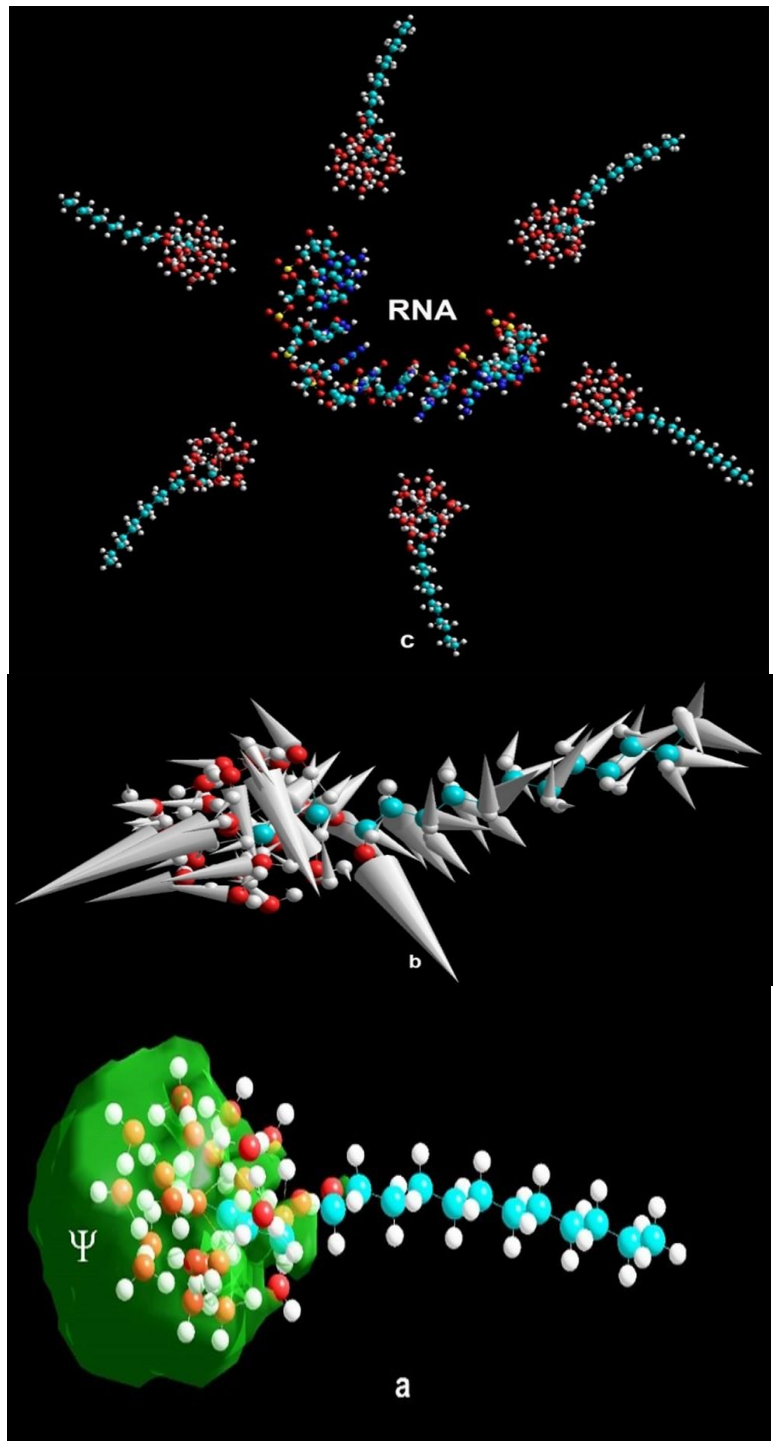

**Fig. 9**

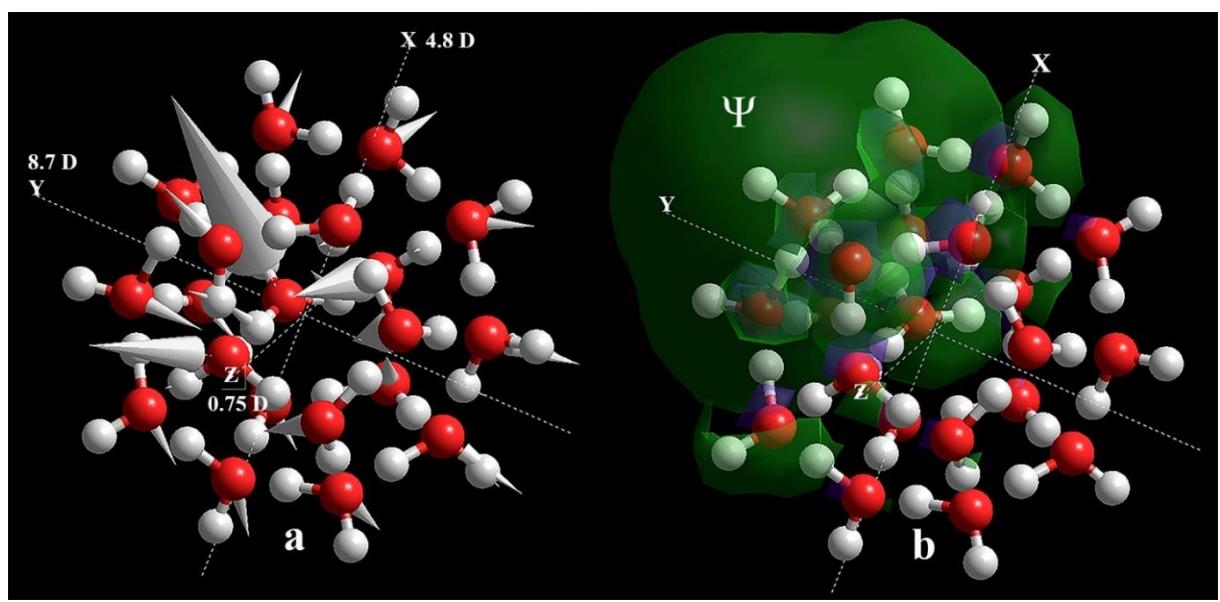

**Fig. 10**